\documentclass[
    aps,
    prd,
    reprint,
    nofootinbib,
    superscriptaddress,
    amsmath,
    amssymb
]{revtex4-2}
\usepackage[utf8]{inputenc}

\usepackage{times}
\usepackage{amsfonts}
\usepackage{amsmath}
\usepackage{amssymb}
\usepackage{natbib}

\usepackage{graphicx}
\usepackage{enumitem}
\usepackage{xcolor}
\usepackage[colorlinks=true,linkcolor=magenta,citecolor=blue]{hyperref}
\usepackage[nameinlink]{cleveref}
\usepackage{soul}
\usepackage[outdir=./]{epstopdf}
\usepackage[normalem]{ulem}
\usepackage{amsmath}
\usepackage{ragged2e}
\usepackage{subcaption}   
\DeclareCaptionJustification{justified}{\justifying}
\usepackage{orcidlink}

\usepackage{newunicodechar}
\newunicodechar{⊙}{\ensuremath{\odot}}

\newcommand{\MBH}{M_{\rm BH}}
\newcommand{\Msun}{M_\odot}
\newcommand{\rplus}{r_+}
\newcommand{\OmH}{\Omega_H}
\newcommand{\Fabs}{F_{\rm abs}}
\newcommand{\GammaA}{\Gamma_{\ell=0}}
\newcommand{\GammaSR}{\Gamma_{\ell=1}}
\newcommand{\Mc}{M_{\rm c}}
\newcommand{\al}{\mu\MBH}

\begin{document}

\definecolor{orange}{rgb}{0.9,0.45,0} 
\definecolor{applegreen}{rgb}{0.055, 0.591, 0.0530}

\title{Dark-to-black super-accretion as a spin-imprinting mechanism for supermassive Kerr black holes}

\author{Saeed Fakhry\texorpdfstring{\orcidlink{0000-0002-6349-8489}}{}}
\email{saeed.fakhry@uv.es}
\affiliation{Departamento de Astronom\'{\i}a y Astrof\'{\i}sica,
Universitat de Val\`encia, Avenida Vicent Andr\'es Estell\'es 19, 46100 Burjassot (Val\`encia), Spain}

\author{Nicolas Sanchis-Gual\texorpdfstring{\orcidlink{0000-0001-5375-7494}}{}}
\affiliation{Departamento de Astronom\'{\i}a y Astrof\'{\i}sica,
Universitat de Val\`encia, Avenida Vicent Andr\'es Estell\'es 19, 46100 Burjassot (Val\`encia), Spain}

\author{Jorge Castelo Mourelle\texorpdfstring{\orcidlink{0000-0002-0074-2608}}{}}
\affiliation{Instituto de Ciencias Nucleares, Universidad Nacional Aut\'onoma de M\'exico, Circuito Exterior C.U., A.P. 70-543, M\'exico D.F. 04510, M\'exico.}

\author{Dar\'io N\'u\~nez Z\'u\~niga\texorpdfstring{\orcidlink{0000-0003-0295-0053}}{}}
\affiliation{Instituto de Ciencias Nucleares, Universidad Nacional Aut\'onoma de M\'exico, Circuito Exterior C.U., A.P. 70-543, M\'exico D.F. 04510, M\'exico.}

\author{Juan Carlos Degollado\texorpdfstring{\orcidlink{0000-0002-8603-5209}}{}}
\affiliation{Instituto de Ciencias F\'{\i}sicas, Universidad Nacional Aut\'onoma de M\'exico, Apdo. Postal 48-3, 62251, Cuernavaca, Morelos, M\'exico}


\begin{abstract} 
The existence of supermassive black holes with masses $M\gtrsim10^9\,M_{\odot}$ and large dimensionless spins $\chi\sim0.9-0.99$ at high redshift remains a challenge to our understanding of the early Universe. In this work, we study the adiabatic co-evolution of a Kerr black hole seed surrounded by two ultralight scalar dark matter clouds occupying different bound states, and show that this configuration allows the black hole to grow into the supermassive mass range while imprinting a characteristic final spin. The evolution proceeds through two stages. During the first stage, a spherical cloud described by the $\ell=0$ mode is completely depleted through a runaway dark-to-black accretion mechanism on a timescale of hundreds of millions of years for boson masses $\mu\sim10^{-18}-10^{-17}\,\mathrm{eV}$. Since the accreted material does not carry angular momentum, the black hole spin is universally driven to $\chi\simeq0$, independently of its initial spin. Throughout this stage, the second cloud, described by the $\ell=m=1$ mode, remains in the superradiant regime with negligible evolution. However, once the first stage is completed, this cloud transitions to the accreting regime, rapidly transferring both mass and angular momentum to the black hole. Starting from $\chi\simeq0$, the black hole spin increases until the evolution self-consistently saturates close to the threshold $\chi_{\rm sat}$, defined by the condition $\Omega_H(\chi_{\rm sat})=\mu$, on an e-folding timescale of thousands of years, orders of magnitude shorter than the first stage. This final saturation spin is largely independent of both the initial black hole spin and the mass of the secondary cloud, providing a spin-imprinting mechanism in which the primordial spin is first erased by spherical accretion and then reset to a value determined only by the boson mass and the final black hole mass. We characterize the saturation spin as a function of the black hole mass and boson mass. 
\end{abstract}

\keywords{ultralight dark matter; black hole superradiance; Kerr black holes;
          dark-to-black accretion; spin evolution; bosonic clouds}

\maketitle


\section{Introduction}
\label{sec:intro}
The discovery of quasars at redshifts $z \gtrsim 6$-$10$ harbouring black holes (BHs) with masses $\MBH \gtrsim 10^9\,\Msun$ represents one of the most acute tensions between observation and theory in high-energy astrophysics~\cite{2006AJ....131.1203F,2011Natur.474..616M,2015Natur.518..512W, 2018Natur.553..473B,2021ApJ...907L...1W,2021ApJ...908L..33Y}. Within the standard $\Lambda$CDM framework, a BH that forms at $z \sim 30$ and accretes continuously at the Eddington rate with a radiative efficiency of ten percent reaches $10^9\,\Msun$ by $z \sim 7$ only if it starts from a seed of at least $10^4\,\Msun$~\cite{2010A&ARv..18..279V, 2020ARA&A..58...27I}. Yet Population~III stellar remnants, the most abundant and least fine-tuned seed channel~\cite{2001ApJ...551L..27M,2004ApJ...604..484M, 2011Sci...334..948T}, reach near a few hundred solar masses, requiring unlikely uninterrupted accretion at or above the Eddington limit for hundreds of millions of years. Heavy seeds formed by the direct collapse of metal-free gas clouds~\cite{1994ApJ...432...52L,2006MNRAS.370..289B,2010MNRAS.402.1249S} can bypass this bottleneck, but require environments of near-zero metallicity bathed in strong Lyman-Werner radiation~\cite{2017MNRAS.469.3329W}, conditions whose cosmic abundance is poorly constrained. Nuclear-star-cluster channels involving runaway tidal disruptions and stellar collisions~\cite{2004Natur.428..724P,2010ARA&A..48..339B} face feedback
barriers that are difficult to overcome at the centres of low-mass halos.
The high-redshift BH mass function is therefore a sharp discriminant
between formation scenarios, and its implications are not yet
resolved~\cite{2020ARA&A..58...27I}.

A fundamentally different growth avenue opens if a fraction of the dark
matter is composed of ultralight bosons with masses
$\mu \sim 10^{-22}$-$10^{-10}$~eV. On astrophysical scales such fields
behave as classical waves rather than particle streams, and their
macroscopic Compton wavelength allows them to reproduce the successes of
cold dark matter on large scales while departing from it below a
characteristic Jeans-like scale set by $\mu$. For $\mu\sim10^{-22}$~eV,
in particular, this ``fuzzy'' dark matter suppresses small-scale power and
produces solitonic, cored halo profiles that alleviate long-standing
small-scale tensions of the cold-dark-matter
paradigm~\cite{Matos:1998vk,2000PhRvL..85.1158H,2017PhRvD..95d3541H}, while remaining
statistically indistinguishable from $\Lambda$CDM on the scales probed by
the cosmic microwave background and the large-scale
matter power spectrum. Bosonic fields with such small masses are not an
ad hoc addition to the dark sector: pseudoscalars spanning many decades in
mass arise generically in string compactifications~\cite{2010PhRvD..81l3530A,
2006JHEP...06..051S}, and the QCD axion itself~\cite{1977PhRvL..38.1440P,
1977PhRvD..16.1791P,1978PhRvL..40..223W,1978PhRvL..40..279Wi}, introduced on
entirely independent grounds to resolve the strong-$CP$ problem, sits only
a few decades away in mass from the fuzzy dark matter window. The ultralight sector is a compelling dark matter candidate motivated by both its cosmological and particle-physics aspects. 

When the boson's Compton wavelength $\lambda_C = 1/\mu$ is much larger
than the gravitational radius of a BH, $\lambda_C \gg 2\MBH$, the scalar
field bounds to the BH forming a \emph{gravitational atom}: a hydrogen-like spectrum of quasi-bound states whose occupation numbers grow or decay on timescales controlled by the dimensionless gravitational coupling $\mu\MBH$, which plays the role of a fine-structure constant~\cite{2019JCAP...12..006B}, see also \cite{Alcubierre:2025zus} for an astrophysical discussion on these objects. The analogy is precise in the weak-coupling regime $\al\ll1$: the energy eigenvalues follow a Bohr-like sequence, the real part of the mode frequency approaches $\mu$, and the imaginary part encodes absorption or superradiant amplification at the horizon~\cite{1980PhRvD..22.2323D,1979AnPhy.118..139Z,2007PhRvD..76h4001D}. Since we assume $\al\ll1$ throughout the mass and coupling range explored in this work, this hydrogenic approach is well supported.

For a Schwarzschild (non-rotating) BH, every quasi-bound state decays
purely by absorption: lacking angular momentum to exchange with the
horizon, the scalar field cloud can only lose mass to the BH, on a period of time
that scales steeply as $\tau_{\rm abs}\propto\mu^{-6}\MBH^{-5}$~\cite{barranco2011black,barranco2012schwarzschild,sanchis2015quasistationarya,sanchis2015quasistationaryb,sanchis2016quasistationary,cardoso2022parasitic}.\footnote{The absorption rates employed in this work are computed using the Detweiler approximation, which assumes a scalar perturbation on a Kerr background. Here we use these rates as instantaneous adiabatic rates evaluated on the evolving Kerr geometry. Although the initial cloud mass may be comparable to or exceed the seed BH mass, we assume that the evolution can be approximated as a sequence of Kerr spacetimes with slowly varying mass and spin. Including the backreaction of the scalar cloud on the spacetime is expected to enhance the accretion rates, thereby potentially accelerating the mechanism, and constitutes a natural extension of the addiabatic approximation used in the present work ~\cite{annulli2020response,cardoso2022parasitic,deCesare:2026fie}.} This is the basis of the dark-to-black super-accretion mechanism recently proposed in Ref.~\cite{2026PhLB..87440251S}, which showed that the runaway absorption of an $\ell=m=0$ scalar cloud by a Schwarzschild seed of $10^2$-$10^5\,\Msun$ can grow the BH to $10^6$-$10^8\,\Msun$
within $\sim10^8$~yr. The mechanism is self-amplifying: because the
absorption rate grows as $\MBH^5$, any incremental mass accelerates further growth, driving the accretion rate to many orders of magnitude above the Eddington limit. Therefore, dark matter is efficiently transformed into BH mass. In~\cite{2026PhLB..87440251S}, it was
established this dark-to-black mechanism for a non-spinning seed; extending it to a Kerr seed, and following the fate of its spin, is the purpose of the present work.

For a Kerr (rotating) BH the picture is qualitatively richer, because
modes with azimuthal number $m\geq1$ can instead be
\emph{superradiantly amplified}, extracting rotational energy and angular
momentum from the BH whenever their frequency lies below the
superradiance threshold, $\omega_R<m\Omega_H$, with $\Omega_H$ the
horizon angular velocity~\cite{1971JETPL..14..180Z,1972JETP...35.1085Z,
1973JETP...37...28S,1972Natur.238..211P,east2014black,2015LNP...906.....B}. In this regime the cloud
grows exponentially at the expense of the BH spin, and the
process self-terminates once the mode frequency matches the horizon
frequency, $\omega_R=m\Omega_H$. 

For a genuinely complex scalar field,
the case considered throughout this work, the end state of this
process is not a radiating transient but an exactly stationary
configuration: a Kerr BH endowed with ``synchronized" scalar
hair, whose helical Killing vector locks the field's phase rotation to
the horizon's rotation. The composite spacetime is stationary,
axisymmetric, and the cloud emits no gravitational
radiation~\cite{Herdeiro:2014goa,Herdeiro:2015waf} (a real, self-gravitating
field condensate would instead possess a slowly time-varying quadrupole and radiate weakly, but this is not the regime relevant here). A structurally similar mechanism, in which a Kerr BH acquires synchronized scalar hair by absorbing both the mass and orbital angular momentum of its bosonic environment, has been shown to operate in the merger of binary boson stars~\cite{Sanchis-Gual:2020mzb}, lending further
support to accretion as a generic spin-imprinting channel for
horizons embedded in a bosonic medium. The superradiant instability itself has been characterized in the scalar case at leading perturbative order~\cite{1980PhRvD..22.2323D,
1979AnPhy.118..139Z,2007PhRvD..76h4001D,2013PhRvD..87l4026D} and in the
time domain~\cite{2012PhRvD..86f4036W,2018PhRvL.121m1104E}, in the charge case~\cite{sanchis2016explosion,bosch2016nonlinear}, and for
vector fields both analytically~\cite{2012PhRvD..86f4019P,
2017PhRvD..96c5019B} and in full nonlinear regime through numerical-relativity simulations~\cite{2017PhRvL.119d1101E}. 

Independently of the mechanism responsible for setting it, BH
spin is now an observable quantity. X-ray reflection spectroscopy, which
fits the relativistically broadened iron K$\alpha$ emission line from the
innermost accretion disk, yields spin estimates for a growing sample of
active galactic nuclei (AGN)~\cite{2014SSRv..183..277R,2021NatAs...5..133R}, complemented by continuum fitting of the thermal disk spectrum~\cite{2006ApJ...636L.113S,2014SSRv..183..295M}. A compilation of these measurements indicates that a large fraction of supermassive BHs are rapidly spinning, with $\chi\gtrsim0.5$, although systematic uncertainties associated with the reflection modelling remain substantial~\cite{2014SSRv..183..277R,2021NatAs...5..133R,2021SSRv..217...65B}. These data provide the observational benchmark against which we compare
our predicted saturation spins in Sec.~\ref{sec:obs}.

In the present work, we extend the dark-to-black scenario of Ref.~\cite{2026PhLB..87440251S} to a spinning seed, proposing it as a  spin-imprinting mechanism for rotating BHs. In this regard, this work is organised as follows. In Section~\ref{sec:setup}, we establish the physical setup: the Kerr geometry, the quasi-bound-state spectrum, and the signed absorption and superradiant rates for the $\ell=0$,  $\ell=1$, and $\ell=2$ modes. In Section~\ref{sec:twophase}, we present the full two-phase sequential dynamics that constitute the core mechanism of this work. Moreover, in Section~\ref{sec:phase2standalone}, we analyze Phase~2 as an independent problem, isolating the universal attractor behaviour of the saturation spin. In Section~\ref{sec:saturation}, we explore the timescales of the superradiant unstable higher $m$ modes ($m>1$)~\cite{ganchev2018scalar,Degollado:2018ypf}, focusing on the $\ell=m=2$ case, and the comparison with AGN spin measurements. Finally, in Section~\ref{sec:conclusions}, we summarize our results and outline directions for future work. Throughout this paper we use geometric units $G=c=\hbar=1$ and quote boson masses in electron Volts,~eVs.

\section{Dark-to-black super-accretion}
\label{sec:setup}

In quantum-mechanical language, an ultralight scalar field around a BH can form a hierarchy of quasi-bound states labelled by three quantum numbers with a structure reminiscent of the hydrogen atom: the principal number $n$, the orbital number $\ell$, and the azimuthal number $m$. The scenario developed in this paper considers the coupled evolution of a single Kerr BH and a scalar field composed of two quasi-bound modes with different angular quantum numbers. Because of the linearity of the Klein-Gordon equation, the scalar field is treated as a superposition of these modes, whose evolution is assumed to be independent throughout this work. The quasi-bound states are described within the test-field approximation~\cite{barranco2011black,barranco2012schwarzschild}. Since the coupling of each mode to the BH horizon depends sensitively on the angular momentum it carries, the two clouds interact with the BH in qualitatively different ways: the $\ell=0$ mode is always absorbed, whereas the $\ell=m=1$ mode can be either absorbed or superradiantly amplified depending on the instantaneous BH spin. The underlying physical mechanism is described below. 

Before presenting the dynamical analysis, we briefly discuss possible astrophysical scenarios leading to a BH surrounded by two quasi-bound clouds carrying different angular momenta. Although a detailed formation model lies beyond the scope of this work, several plausible mechanisms may give rise to such a configuration. One possibility is a primordial black hole forming within a virialized ultralight dark matter halo, where it is naturally immersed in a scalar field environment associated with the ambient dark matter distribution~\cite{2017PhRvD..95d3541H, 2020ARNPS..70..355C,guzman2020gravitational,sanchis2021multifield}. Due to the linearity of the Klein-Gordon equation in the test-field limit, the scalar field is generically described by a superposition of quasi-bound states rather than a single mode, allowing gravitational capture to populate multiple states simultaneously~\cite{2019JCAP...12..006B,guzman2022possible,bernal2025natural}. As the halo evolves, the relative occupation of these states may change through continued gravitational capture and accretion of the surrounding scalar field. Alternatively, the multistate configuration may represent the remnant of an earlier superradiant or dark-to-black accretion episode, in which one mode has been partially depleted while another remains populated~\cite{2018PhRvL.121m1104E,2017PhRvL.119d1101E}. Throughout this work, we adopt such a multistate configuration as a physically motivated initial condition. We first recall, in Sec.~\ref{sec:schw}, the non-rotating limit in which the dark-to-black super-accretion mechanism was originally introduced~\cite{2026PhLB..87440251S}. In Sec.~\ref{sec:kerr}, we review the Kerr geometry and introduce the superradiance condition, before deriving the absorption and superradiant rates governing the evolution of the two quasi-bound states considered in our two-phase scenario.

\subsection{Schwarzschild metric}
\label{sec:schw}

The starting point of the dark-to-black scenario is the absorption of an
ultralight scalar cloud by a non-rotating BH.
In the Schwarzschild background, the massive Klein-Gordon equation assuming no self-interactions for the field 
$(\nabla^\alpha\nabla_\alpha - \mu^2)\Psi = 0$ admits quasi-bound-state
solutions of the form $\Psi(t,r,\theta,\phi) = e^{i(m\phi-\omega t)}\psi_\ell(r,\theta)$, where $m$ is the azimuthal index or winding number, $\omega$ is the complex frequency of the field, and the $\theta$ dependence is given by the $\ell$ orbital angular number of the spherical harmonics. The normalizable solutions are purely ingoing at the horizon and exponentially decaying at spatial infinity. The complex frequency $\omega = \omega_{\rm R} + i\omega_{\rm I}$ encodes both the oscillation frequency of the state and the rate at which the scalar field is absorbed by the BH. In the weak-coupling regime $\mu\MBH \ll 1$, matched asymptotic expansion of the radial equation across the near-horizon and
far-field regions reproduces a hydrogen-like spectrum for the real part of
the frequency~\cite{1980PhRvD..22.2323D,1979AnPhy.118..139Z},
\begin{equation}
  \omega_{\rm R} = \mu\left[1 - \frac{(\al)^2}{2n^2} + \mathcal{O}\!\left((\al)^4\right)\right],
  \label{eq:omega_R}
\end{equation}
where $n$ is the principal quantum number given by $n\geq\ell+1$, and $\al$ plays the role of the
fine-structure constant of this ``gravitational atom". The imaginary part
$\omega_I<0$ quantifies the exponential decay of the field across the horizon; for
the fundamental $\ell=m=0$, $n=1$ mode, the leading-order result
is~\cite{1980PhRvD..22.2323D,2023PhRvD.107d4070A,2026PhLB..87440251S}
\begin{equation}
  \omega_{\rm I} = -8\mu^6\MBH^5,
  \label{eq:omega_I_schw}
\end{equation}
giving an absorption rate $\GammaA^{\rm Schw} \equiv 2\omega_{\rm I} = -16\mu^6\MBH^5$.
Two features of this result drive the dark-to-black mechanism.
First, the rate depends on $\MBH$ through the steep power $\MBH^5$
(equivalently $(\al)^5$ at fixed $\mu$): a BH that has grown even modestly absorbs at a much higher rate than its seed. Second, the process is entirely dissipative, there is no channel by which the
scalar field can extract energy from a non-rotating BH, thus
$\GammaA^{\rm Schw}$ remains negative at all times. 
The coupled evolution of the BH mass $\MBH$ and the scalar cloud mass
$\Mc$ under this rate follows immediately from energy conservation applied to the two-body system,
\begin{equation}
  \dot{M}_{\rm BH} = -\GammaA^{\rm Schw}\,\Mc, \qquad
  \dot{M}_{c}  = \GammaA^{\rm Schw}\,\Mc,
  \label{eq:schw_evo}
\end{equation}
which conserves $\MBH + \Mc$ exactly at every instant, as it must for a
process in which the cloud's rest-mass energy is completely transferred across the horizon. Equations~\eqref{eq:omega_I_schw} and~\eqref{eq:schw_evo} close a positive-feedback loop: as the BH absorbs the cloud mass its own mass increases, which raises $\GammaA^{\rm Schw}\propto\MBH^5$ further, which in turn accelerates the absorption. The result is a runaway, attractor-like growth that
self-terminates only once the cloud reservoir $\Mc$ is exhausted. Once
$\al$ grows beyond $\al\sim0.01$-$0.1$, the instantaneous accretion rate implied by Eq.~\eqref{eq:schw_evo} exceeds the Eddington limit by many orders of magnitude, providing a super-Eddington channel for BH mass growth~\cite{2026PhLB..87440251S}. We will generalize to spinning seeds in the remainder of this section, but this analysis will constitute what we have called Phase~1 of the two-phase scenario discussed in Sec.~\ref{sec:twophase}.

\subsection{Kerr geometry and superradiance}
\label{sec:kerr}

The analysis of the Schwarzschild case in Sec.~\ref{sec:schw} assumes zero spin, but
any realistic seed BH formed through the collapse of baryonic matter will generically carry angular momentum, and the
absorption of a spherically symmetric $\ell=m=0$ cloud alone cannot account for the large dimensionless spins
$\chi=J/{M_{BH}}^2\sim0.9$-$0.99$ inferred for high-redshift quasars~\cite{2025IJMPD..3450046P}. Understanding how spin can be introduced in this problem therefore requires the study of modes carrying non-zero angular momentum, ($m \neq 0)$, and of the full Kerr geometry, which in Boyer-Lindquist coordinates reads
\begin{align}
  ds^2 &= -\left(1-\frac{2\MBH r}{\Sigma}\right)dt^2
           -\frac{4\MBH ar\sin^2\!\theta}{\Sigma}\,dt\,d\phi
           +\frac{\Sigma}{\Delta}\,dr^2 \nonumber\\
       &\quad
           +\Sigma\,d\theta^2
           +\left(r^2+a^2+\frac{2\MBH a^2 r\sin^2\!\theta}{\Sigma}\right)
            \sin^2\!\theta\,d\phi^2,
  \label{eq:Kerr}
\end{align}
where $\Sigma \equiv r^2 + a^2\cos^2\!\theta$,
$\Delta \equiv r^2 - 2\MBH r + a^2$, and $a \equiv \chi\MBH$ is the
specific angular momentum, with $\chi \in [0,1)$. The event horizon lies at the largest root of
$\Delta(r)=0$,
\begin{equation}
  \rplus = \MBH\!\left(1 + \sqrt{1-\chi^2}\right),
  \label{eq:rplus}
\end{equation}
and, being a Killing horizon, co-rotates rigidly with angular velocity
$\OmH = -g_{t\phi}/g_{\phi\phi}\big|_{r=\rplus}$, which for
metric~\eqref{eq:Kerr} evaluates to the standard expression
$\OmH = a/(\rplus^2+a^2)$. Using the horizon condition
$\Delta(\rplus)=0$, i.e.\ $\rplus^2+a^2 = 2\MBH\rplus$, together with
$a=\chi\MBH$, this can be written compactly in terms of $\chi$ and
$\MBH$ alone,
\begin{equation}
  \OmH = \frac{a}{2\MBH\rplus}=\frac{\chi}{2\MBH\!\left(1+\sqrt{1-\chi^2}\right)}.
  \label{eq:OmH}
\end{equation}
Equation~\eqref{eq:OmH} shows that $\OmH$ is a strictly increasing
function of $\chi$, bounded from above by $1/(2\MBH)$ in the extremal limit
$\chi\to1$; this upper bound plays a central role in determining the
saturation spin studied in Sec.~\ref{sec:saturation}, since it sets a
maximum frame-dragging frequency that the BH can offer to a cloud of
given $\mu$.

A massive scalar field propagating on the Kerr background admits
separable solutions
$\Psi = e^{i(m\phi-\omega t)}S_{\ell m}(\theta)R_{\ell m}(r)$, where
$S_{\ell m}$ are spheroidal harmonics labelled by the orbital and
azimuthal quantum numbers $\ell$ and $m$~\cite{2024PhRvD.110l4064C}. 
The modes are thus characterized by the quantum numbers $\ell$, $m$ and $n$. We refer to these states as ($\ell$, $m$, $n$).
The energy flux carried by such a mode across the horizon can be obtained from the Wronskian of the radial equation evaluated between the horizon and infinity, and is proportional to the combination $(\omega_{\rm R}-m\OmH)$~\cite{1980PhRvD..22.2323D,2015LNP...906.....B}. 
This sign reversal is the defining feature of superradiant scattering: because the boundary condition at the horizon forces purely ingoing flux in the locally co-rotating frame, a mode with
\begin{equation}
  \omega_{\rm R} < m\OmH
  \label{eq:SR_condition}
\end{equation}
must be reflected back to infinity with an amplitude \emph{larger} than
its incident value, extracting rotational energy and angular momentum from
the BH in a wave-mechanical analogue of the classical Penrose
process~\cite{1971JETPL..14..180Z,1972JETP...35.1085Z,1973JETP...37...28S,
1972Natur.238..211P,2015LNP...906.....B}. A mode with
$\omega_{\rm R}>m\OmH$ is damped: it deposits both energy and angular momentum into the horizon and is simply absorbed. At leading order in the weak-coupling expansion of Eq.~\eqref{eq:omega_R},
$\omega_{\rm R}\to\mu$ as $\al\to0$, thus the superradiant
condition~\eqref{eq:SR_condition} reduces, up to fractional corrections of
order $(\al)^2/2n^2$, to the familiar threshold $\mu<m\OmH$. These corrections are controlled throughout our analysis: for the $(1,1,2)$ state that dominates Phase~2, they amount to at most a few percent over the range $\al\lesssim0.5$ realized in our numerical integrations, and their quantitative effect on the saturation spin is computed explicitly in Sec.~\ref{sec:chi_sat}. The Kerr problem is thus qualitatively richer than its Schwarzschild counterpart: modes with
$m\geq1$ can be \emph{either} absorbed or amplified depending on the instantaneous spin state of the BH, so that the signed rate,
rather than a rate of fixed sign as in Eq.~\eqref{eq:omega_I_schw}, is
the key quantity governing the co-evolution developed below.

\subsection{Absorption and superradiant rates in the Kerr spacetime}
\label{sec:rates}

The two phases of the dark-to-black mechanism studied here involve different azimuthal modes of the scalar field, and before writing the rate formulas it is useful to describe the physical nature of each cloud and its coupling to the BH.

The gravitational potential of the BH plays the role of the Coulomb potential, while the gravitational coupling $\mu\MBH$ plays the role of the fine-structure constant~\cite{2019JCAP...12..006B}, with states localized closer to the horizon for smaller $n$ and $\ell$. The three dominant states relevant to our scenario are summarized in Table~\ref{tab:modes}.

\begin{table}[t]
\caption{The three quasi-bound states entering the dark-to-black scenario,
the angular momentum carried per quantum, and the leading-order scaling of their interaction rate with the BH, where $ \mu\MBH \ll 1$.}
\label{tab:modes}
\begin{ruledtabular}
\begin{tabular}{lcc}
 State $(\ell,m,n)$ & $\hbar$ per quantum & Rate \\
\hline
 $(0,0,1)$  & $0$ & $\GammaA\propto (\al)^{5}$ \\
 $(1,1,2)$  & $1$ & $\GammaSR\propto (\al)^{9}$ \\
 $(2,2,3)$  & $2$ & $\Gamma_{\ell=2}\propto (\al)^{14}$ \\
\end{tabular}
\end{ruledtabular}
\end{table}

The ($\ell, m, n)=(0,0,1)$ state is the lowest spherically symmetric member of the quasi-bound spectrum. It is centered around $r\sim1/\mu$, and carries no azimuthal angular momentum ($m=0$). Consequently the superradiance condition, $\omega_R<m\OmH$, can never be satisfied for this mode, and it interacts with the BH solely through horizon absorption. Each absorbed quantum increases the BH mass by $\omega_R\simeq\mu$ without transferring any angular momentum. For a Schwarzschild BH, the absorption cross section
scales as $\sigma_{\rm Schw}\propto\MBH^2$. However, in the Kerr case it is reduced by the factor $\Fabs(\chi)=\sqrt{1-\chi^2}$, reflecting the shrinking absorption probability of a horizon with reduced surface gravity~\cite{1973CMaPh..31..161B}. The corresponding absorption rate is
therefore
\begin{equation}
  \GammaA(\mu,\MBH,\chi) = -16\,\mu^6\MBH^5\sqrt{1-\chi^2},
  \label{eq:Gamma_l0}
\end{equation}
which reduces to the Schwarzschild result~\eqref{eq:omega_I_schw} at
$\chi=0$ and vanishes in the extremal limit $\chi\to1$, where the horizon area shrinks to its minimal value at fixed $\MBH$. Because $\GammaA$ never changes sign, this mode is responsible for effectively erasing, rather than providing, spin. As we show in Sec.~\ref{sec:phase1}, its absorption drives $\chi\to0$ universally, independently of the initial spin $\chi_0$, while simultaneously fuelling the super-Eddington mass growth of Phase~1.

The $(1,1,2)$ state is the first quasi-bound mode that carries angular momentum. Unlike the spherical $(0,0,1)$ state, its density is concentrated around the equatorial plane in a torus-like configuration, with each scalar quantum carrying one unit of azimuthal angular momentum. This mode can either extract or transfer angular momentum depending on the BH spin. The horizon angular velocity $\Omega_H$ determines the boundary
between these two regimes: when $\Omega_H > \omega_R/m \simeq \mu$,
the field mode rotates more slowly than the horizon and is
superradiantly amplified, extracting rotational energy and angular
momentum from the BH; when $\Omega_H < \mu$, the horizon rotates
more slowly than the field and the amplification condition is no
longer met, so the cloud is gradually absorbed, transferring both mass \emph{and} angular momentum to the BH. This behavior, absent for the $\ell=0$ mode, allows the $\ell=m=1$ cloud to act as a spin-imprinting reservoir rather than a purely dissipative one. The interaction is described by the signed rate obtained from matched asymptotic solutions of the Teukolsky equation~\cite{1980PhRvD..22.2323D},
\begin{eqnarray}
  \GammaSR(\mu,\MBH,\chi)
  = \frac{1}{24}\,\frac{\rplus}{\MBH}
    \left[1 - 4\rplus\mu\bigl(\OmH - \mu\bigr)\right]\nonumber\\
   \times \bigl(\OmH - \mu\bigr)\,(\mu\MBH)^9.
  \label{eq:Gamma_l1}
\end{eqnarray}
The sign of $\GammaSR$ tracks the sign of $(\OmH-\mu)$: $\GammaSR<0$ for
$\OmH<\mu$, corresponding to a decaying cloud, whereas $\GammaSR>0$ for
$\OmH>\mu$, indicating superradiant growth. The transition between the
two regimes occurs at $\GammaSR=0$, i.e.\ at
\begin{equation}
  \OmH(\chi_{\rm sat}) = \mu,
  \label{eq:chi_SR_def}
\end{equation}
which defines the saturation spin $\chi_{\rm sat}^{(1)}$: the unique
spin at which the $\ell=m=1$ mode is neither amplified nor damped.
Substituting Eq.~\eqref{eq:OmH} and solving, this condition yields the
closed-form expression
\begin{equation}
  \frac{\chi_{\rm sat}^{(1)}}{2\!\left(1+\sqrt{1-[\chi_{\rm sat}^{(1)}]^2}
  \right)} = \al,
  \label{eq:chi_sat_general}
\end{equation}
which admits a real solution if and only if $\mu M_{\rm BH}<1/2$.
Above this coupling, the boson frequency exceeds the maximum horizon frequency attainable at extremality, $m\Omega_H^{\rm max} = m/(2M_{\rm BH})$, so the synchronization condition cannot be satisfied within the approximation $\omega_R \simeq \mu$ adopted throughout this work. However, the existence of quasi-bound states is not affected by this limitation, as they persist beyond this coupling. Instead, the limitation arises because the leading-order frequency relation in Eq.~\eqref{eq:omega_R} ceases to be accurate as $\mu M_{\rm BH} \to m/2$. The generalization to arbitrary azimuthal number $m$ and the resulting mass cutoff are derived in Sec.~\ref{sec:chi_sat}.

For the fiducial values $\mu=1.73\times10^{-17}\,\mathrm{eV}$ and $\MBH=10^6\,\Msun$ adopted in Sec.~\ref{sec:twophase}, the corresponding saturation spin is $\chi_{\rm sat}^{(1)}\simeq 0.517$, which serves as the reference value for the Phase~2 analysis in Sec.~\ref{sec:phase2}. For brevity, whenever the $\ell=m=1$ channel is the only relevant channel, as is the case throughout Sec.~\ref{sec:twophase}, we follow standard practice and drop the superscript, writing simply $\chi_{\rm sat}\equiv\chi_{\rm sat}^{(1)}$; the superscript is restored below and in Sec.~\ref{sec:saturation} whenever more than one mode is discussed simultaneously.
 
The $(2,2,3)$ state has a more extended torus-like distribution around the equatorial plane. Its interaction with the BH is governed by the same superradiant mechanism as the $(1,1,2)$ mode, except that the amplification condition becomes $2\Omega_H>\mu$. The corresponding saturation spin, defined by $2\Omega_H(\chi_{\rm sat}^{(2)})=\mu$, is therefore lower than $\chi_{\rm sat}^{(1)}$ for the same value of $\mu M_{\rm BH}$. The corresponding growth rate, derived in Appendix~\ref{app:gamma_l2}, is
\begin{equation}
  \Gamma_{\ell=2}(\mu,\MBH,\chi)
  = \frac{1}{23040}
    \left(\frac{\rplus}{\MBH}\right)^{5}
    \frac{2\OmH-\mu}{\MBH}
    (\mu\MBH)^{14},
  \label{eq:Gamma_l2}
\end{equation}
with the same sign convention as $\GammaSR$. Comparing
Eqs.~\eqref{eq:Gamma_l1} and~\eqref{eq:Gamma_l2},
\begin{equation}
  \frac{|\Gamma_{\ell=2}|}{|\GammaSR|} \sim (\al)^{5} \ll 1,
  \qquad \al<0.5,
  \label{eq:rate_ratio}
\end{equation}
shows the higher power of $\al$ with which the $\ell=2$ rate scales relative to $\ell=1$ [cf.\ Table~\ref{tab:modes}]. Physically, this suppression originates from the larger centrifugal barrier of the $\ell=2$ state, which pushes its wavefunction further from the horizon and reduces its overlap with the near-horizon region where superradiant amplification is sourced. In this work, however, we consider the $\ell=m=2$ mode only as a consistency check of the two-phase scenario. We compute its superradiant growth timescale to compare it with the age of the Universe to determine whether the remnant $\ell=m=1$ cloud can trigger a subsequent $\ell=m=2$ superradiant instability.

Finally, we note that Eqs.~\eqref{eq:schw_evo}, together with the analogous evolution equations for $\MBH$ and $\chi$ derived from Eqs.~\eqref{eq:Gamma_l0} and~\eqref{eq:Gamma_l2}, treat $\GammaA$, $\GammaSR$, and $\Gamma_{\ell=2}$ as instantaneous rates evaluated on the evolving Kerr background. This quasi-adiabatic approximation assumes that the BH parameters evolve on timescales much longer than the oscillation period of the scalar cloud, allowing the field to continuously adjust through a sequence of quasi-bound states. We verify this assumption a posteriori in Secs.~\ref{sec:phase1} and~\ref{sec:phase2} by showing that the characteristic evolution timescales of both $\MBH$ and $\chi$ remain much longer than $\mu^{-1}$ throughout the evolution.

The absorption and superradiant rates employed throughout this work are obtained within the Detweiler approximation, which assumes a test scalar field on a Kerr background and is formally valid in the hydrogenic regime $\alpha\equiv\mu M_{\rm BH}\ll1$. In the scenarios considered here, $\alpha$ remains below $\mathcal{O}(0.5)$ throughout the evolution, so our calculations should be regarded as an extrapolation of the perturbative rates towards the upper end of their range of applicability. A more accurate treatment, including the gravitational backreaction of the scalar cloud and numerical quasi-bound-state frequencies and growth rates, is left for future work.

\begin{figure*}[!ht]
  \centering  \includegraphics[width=0.93\textwidth]{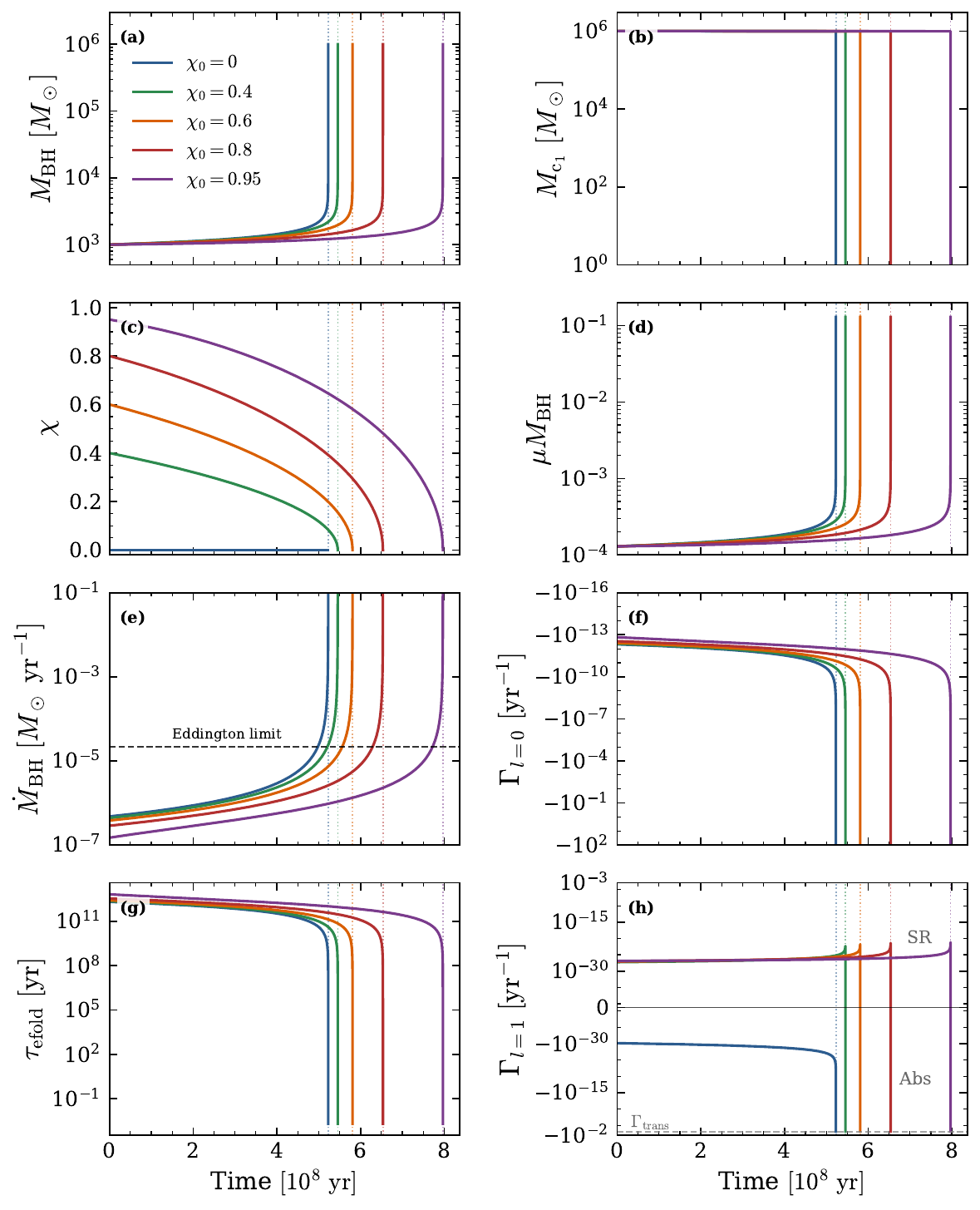}
    \caption{
    Phase~1 co-evolution for $\mu = 1.73\times10^{-17}$~eV,
    $\MBH^{(0)} = 10^3\,\Msun$, $M_{c_1} = 10^6\,\Msun$, and
    five initial spins $\chi_0 \in \{0, 0.4, 0.6, 0.8, 0.95\}$.
    {\bf (a)} BH mass $\MBH$; {\bf (b)} cloud mass $M_{c_1}$;
    {\bf (c)} spin $\chi$; {\bf (d)} dimensionless coupling $\mu\MBH$;
    {\bf (e)} accretion rate $\dot{M}_{\rm BH}$, with the Eddington limit
    shown as a dashed horizontal line; {\bf (f)} absorption rate $\GammaA$;
    {\bf (g)} Phase~1 e-folding time $\tau_{\rm efold} = |\GammaA|^{-1}$;
    {\bf (h)} signed superradiant rate $\GammaSR$ on a symmetric-log axis, with the superradiance (SR) and absorption (Abs) regimes and the transition rate $\Gamma_{\rm trans}$ indicated. Dotted vertical lines in every panel mark the end of phase 1 $t_{\rm end}(\chi_0)$ for each of the five cases. All trajectories converge to $(\MBH,\chi)\approx(10^6\,\Msun,\,0)$
    at the Phase~1 endpoint irrespective of $\chi_0$.
  }
  \label{fig:phase1}
\end{figure*}

\section{Sequential two-phase evolution}
\label{sec:twophase}

\subsection{Parameter choices and numerical setup}
\label{sec:params}
As a fiducial and internally consistent setup, we initially fix the boson mass to $\mu = 1.73\times10^{-17}$~eV, the BH seed mass to $\MBH^{(0)} = 10^3\,\Msun$, and the first-cloud mass to $M_{c_1} = 10^6\,\Msun$. These choices define our reference model and allow us to isolate the impact of the remaining parameters. Later, we will relax these assumptions and extend the analysis over a broad range of boson and BH masses to demonstrate the generality of our results.

To explore the role of the initial spin we consider five values $\chi_0 \in \{0, 0.4, 0.6, 0.8, 0.95\}$, spanning the full range from a Schwarzschild seed to a near-extremal configuration. For Phase~2 we vary the second-cloud mass over $M_{c_2} \in \{10^5,\,5\times10^5,\,10^6,\,5\times10^6\}\,\Msun$ to assess the dependence of the spin-up on the available cloud reservoir.

The equations of motion for both phases are integrated numerically using a stiff Runge-Kutta solver (Radau method) with relative tolerance $10^{-9}$ and absolute tolerance $10^{-30}$. Integration is terminated when the cloud mass falls below $1\,\Msun$, at which point the mass transfer is complete to better than one part in $10^6$ of the initial cloud mass.

\subsection{Phase~1: Spin dilution through cloud absorption}
\label{sec:phase1}

During Phase~1 the system comprises three bodies: the BH, the first
($\ell=m=0$) cloud of mass $M_{c_1}$, and the second ($\ell=m=1$) cloud
of mass $M_{c_2}$.
The complete equations of motion are
\begin{align}
  \dot{M}_{\rm BH} &= -\GammaA\,M_{c_1} - \GammaSR\,M_{c_2},
                                                \label{eq:p1_Mdot_full}\\
  \dot{M}_{c_1} &=  \GammaA\,M_{c_1},               \label{eq:p1_Mcdot_full}\\
  \dot{M}_{c_2} &=  \GammaSR\,M_{c_2},              \label{eq:p1_Mc2dot_full}\\
  \dot{\chi}   &=  \frac{2\chi}{\MBH}
    \left(\GammaA\,M_{c_1} + \GammaSR\,M_{c_2}\right)
    - \frac{\GammaSR\,M_{c_2}}{\mu\MBH^2},          \label{eq:p1_chidot_full}
\end{align}

which exactly conserve the total mass $\MBH + M_{c_1} + M_{c_2}$ at
every instant.
As established quantitatively in the left panel of Fig.~\ref{fig3},
the ratio $|\GammaSR|/\GammaA$ remains negligible over the
entire parameter space relevant to Phase~1, and in practice falls
15 or more orders of magnitude below $\GammaA$ at the couplings
$\al \sim 10^{-4}$-$10^{-2}$ realised during the early evolution.
The $\ell=m=1$ terms in Eqs.~\eqref{eq:p1_Mdot_full}-\eqref{eq:p1_chidot_full}
are therefore entirely negligible throughout Phase~1, and $M_{c_2}$
remains effectively constant.

Our numerical results are summarized in Fig.~\ref{fig:phase1}, where
the left-hand panels track the BH properties (mass, spin, accretion
rate, and e-folding timescale), while the right-hand panels follow the
cloud evolution and the interaction rates.
Together, these results reveal the complete physical sequence driving
the conversion of a dark bosonic cloud into BH mass.

Fig.~\ref{fig:phase1}(a) shows that the BH undergoes an extended
quiescent phase before entering a rapid runaway growth, during which
it gains nearly three orders of magnitude in mass within only a small
fraction of $t_{\rm end}$, 
defined as the time of the
complete depletion of the
cloud and the end of Phase~1.
This behavior arises from the strong dependence of the absorption rate
on the BH mass, $\GammaA\propto\MBH^5$.
As shown in Fig.~\ref{fig:phase1}(f), $|\GammaA|$ increases by
approximately fifteen orders of magnitude over the course of the
evolution, from $\sim10^{-13}\,{\rm yr}^{-1}$ for the initial seed to
$\gtrsim10^2\,{\rm yr}^{-1}$ near $t_{\rm end}$.
The resulting increase in the absorption efficiency leads to the
observed runaway growth without requiring any additional physical
ingredients.

Fig.~\ref{fig:phase1}(d) shows that the dimensionless coupling
$\al=\mu\MBH$ increases from $\simeq1.3\times10^{-4}$ to
$\simeq0.129$ during Phase~1 while remaining within the perturbative
regime $\al\ll1$ assumed in Sec.~\ref{sec:rates}.
Fig.~\ref{fig:phase1}(e) illustrates that the accretion rate exceeds
the Eddington limit by several orders of magnitude over the entire
runaway phase, confirming that the growth mechanism is genuinely
super-Eddington and unconstrained by radiation pressure.
Fig.~\ref{fig:phase1}(g) shows that the corresponding e-folding time
drops from $\gtrsim10^{11}$~yr to less than one year, signalling the
rapid transition from an essentially frozen cloud to an almost
instantaneous collapse on cosmological timescales.

Fig.~\ref{fig:phase1}(b) indicates that the depletion of $M_{c_1}$
precisely tracks the growth of $\MBH$: throughout the evolution
$\MBH+M_{c_1}$ is conserved to the accuracy of the sequential
approximation, confirming that the entire increase in BH mass is
supplied by the first bosonic cloud, which is almost completely
exhausted by $t_{\rm end}$.

A remarkable result of Phase~1 is the universal erasure of the BH
spin.
As shown in Fig.~\ref{fig:phase1}(c), all five trajectories, covering
initial spins $\chi_0\in[0,0.95]$, converge to $\chi\simeq0$ by the
end of Phase~1.
Since the $\ell=m=0$ cloud carries no azimuthal angular momentum,
$J=\chi\MBH^2$ remains constant while the BH mass increases by
approximately three orders of magnitude, driving
$\chi=J/\MBH^2\approx0$ independently of the initial value.

Although the Kerr suppression factor $\Fabs(\chi)=\sqrt{1-\chi^2}$
modifies the evolution rate, our results indicate that it affects only
the \emph{duration} of Phase~1, not its endpoint.
Higher-spin seeds absorb more slowly due to their smaller effective
horizon cross-section, delaying the transition from
$5.23\times10^8$~yr for $\chi_0=0$ to $7.97\times10^8$~yr for
$\chi_0=0.95$.
The initial spin therefore determines when the transition occurs but
leaves no memory in the final state.

Finally, Fig.~\ref{fig:phase1}(h) quantifies the dynamical separation between the two clouds during Phase~1. Rotating BHs with $\chi_0 > \chi_{\rm sat}$ initially satisfy the superradiant condition ($\GammaSR>0$), but the $\ell=m=1$ mode is
completely suppressed relative to the $\ell=0$ absorption throughout:
as shown in the left panel of Fig.~\ref{fig3}, the ratio
$|\GammaSR|/|\GammaA|$ remains below $10^{-3}$ over the entire
parameter space and falls 15 or more orders of magnitude below
$|\GammaA|$ for  $\al\sim10^{-4}$-$10^{-2}$ relevant
during the early evolution.
The $\ell=m=1$ mode therefore contributes at the sub-per-mille level
to the Phase~1 dynamics and can be consistently neglected.
The Phase-1 trajectories overplotted on the left panel of
Fig.~\ref{fig3} confirm this picture directly in the $(\chi,\mu\MBH)$
plane: each coloured curve begins at its initial spin and sweeps toward $(\chi\approx0,\,\mu\MBH\approx0.129)$
entirely within the absorption region where
$|\GammaSR|/|\GammaA|\ll1$.
Only after $\GammaA$ vanishes with the depletion of the first cloud
does the $\ell=m=1$ mode become dynamically dominant.
As the spin falls below $\chi_{\rm sat}$ during Phase~1, $\GammaSR$
crosses zero and enters the absorption regime, marking the onset of
Phase~2.

\subsection{Phase~2: Spin-up through $\ell=m=1$ cloud absorption}
\label{sec:phase2}

As discussed in the previous subsection, within the framework of our analysis, Phase~1 ended with the complete depletion of the $l=m=0$ cloud by the BH, leaving a supermassive BH with mass $\MBH\simeq10^6\,\Msun$ but with almost vanishing spin, $\chi\simeq0$.

\begin{figure*}[!ht]
  \centering
  \includegraphics[width=\textwidth]{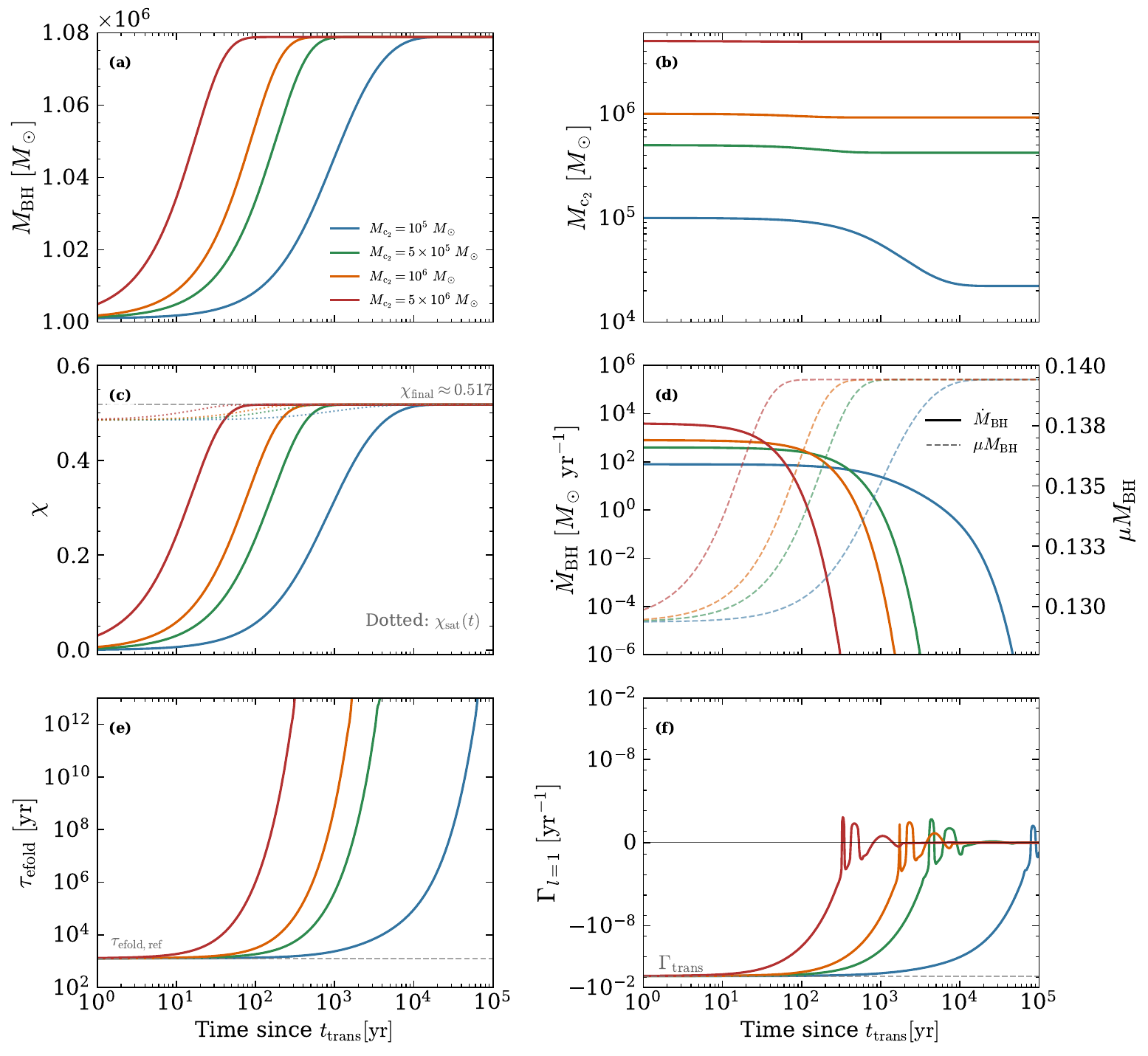}
   \caption{
    Phase~2 co-evolution for $\mu = 1.73\times10^{-17}$~eV,
    $\MBH = 10^6\,\Msun$, $\chi = 0$ at $t_{\rm end}$, and four second-cloud masses. Time is measured from $t_{\rm trans}$, the moment at which the
$\ell=0$ cloud is fully depleted and Phase~2 begins, and 
    $M_{c_2} \in \{10^5,\,5\times10^5,\,10^6,\,5\times10^6\}\,\Msun$.
    {\bf (a)} BH mass in units of $10^6\,\Msun$; {\bf (b)} cloud mass $M_{c_2}$;
    {\bf (c)} spin $\chi$ (solid) with the instantaneous saturation spin
    $\chi_{\rm sat}[\MBH(t)]$ (dotted) and the asymptotic value
    $\chi_{\rm sat}\approx0.517$ (dashed horizontal);
    {\bf (d)} accretion rate $\dot{M}_{\rm BH}$ (solid, left axis) and
    coupling $\mu\MBH$ (dashed, right axis);
    {\bf (e)} Phase~2 e-folding timescale $\tau_{\rm efold}$, with the
    reference value $\tau_{\rm efold, ref}$ shown as a dashed
    horizontal line;
    {\bf (f)} signed $\GammaSR$ on a symmetric-log axis, with the
    transition rate $\Gamma_{\rm trans}$ indicated.
    All trajectories saturate at $\chi_{\rm final}\approx0.517$,
    slightly above $\chi_{\rm sat}(M_{\rm f1})\approx0.485$.
  }
  \label{fig:phase2}
\end{figure*}

We then follow its interaction with the second bosonic cloud carrying $\ell=m=1$ mode, whose dynamics are governed by the signed rate of Eq.~\eqref{eq:Gamma_l1}. The coupled evolution of the system during Phase~2 is now described by the following set of equations:
\begin{align}
  \dot{M}_{BH} &= -\GammaSR\,M_{c_2},                                   \label{eq:p2_Mdot}\\
  \dot{M}_{c_2} &=  \GammaSR\,M_{c_2},                      \label{eq:p2_Mcdot}\\
  \dot{\chi} &=  \GammaSR\,M_{c_2}
    \left[\frac{2\chi}{\MBH}-\frac{1}{\mu\MBH^2}\right], \label{eq:p2_chidot}
\end{align}
where the second term in Eq.~\eqref{eq:p2_chidot} accounts for the angular momentum carried by the cloud. Since Phase~1 leaves the BH with $\chi\simeq0$, the horizon angular velocity initially satisfies $\Omega_H\ll\mu$, implying $\GammaSR<0$. The evolution therefore begins entirely in the absorption regime.

Evaluating the rate at the onset of Phase~2 gives the characteristic
e-folding time
\begin{equation}
  \tau_{\rm efold}
  \equiv
  \frac{1}{|\GammaSR|}
  \bigg|_{\MBH=10^6\Msun,\;\chi=0}
  \approx1.3\times10^3\,{\rm yr},
  \label{eq:tau_efold}
\end{equation}
which is four to five orders of magnitude shorter than $t_{\rm end}$. This shorter timescale reflects the fact that the system enters Phase~2 with an already large value of $\alpha$, leading to a substantially higher absorption rate. Our results therefore show that, although Phase~2 determines the final BH spin, its duration is effectively negligible compared to the much longer evolution of Phase~1.

In Fig.~\ref{fig:phase2}, we present the complete evolution of the system during Phase~2, highlighting four main features. First, the absorption of the $\ell=m=1$ cloud drives the BH spin monotonically toward the saturation value $\chi_{\rm sat}$ defined by Eq.~\eqref{eq:chi_SR_def}. As $\chi$ approaches this value from below, $|\Gamma_{\rm SR}|$ decreases smoothly to zero, naturally suppressing the spin-up. At the same time, the continued growth of the BH mass increases the coupling $\mu\MBH$, causing the saturation spin to shift gradually toward larger values. Figure~\ref{fig:phase2}(c) shows this instantaneous equilibrium as dotted curves, while the numerical solutions continuously track the moving attractor from below. For all cloud masses considered, the system reaches this attractor within $\sim10^4$~yr and converges to the common saturation value $\chi_{\rm final}\simeq0.517$. This value is slightly larger than $\chi_{\rm sat}(M_{\rm f1})\simeq0.485$, evaluated at the end of Phase~1, because the BH continues to accrete mass throughout Phase~2.

As shown in Fig.~\ref{fig:phase2}(a), the BH mass increases only modestly during this stage, gaining at most $\sim8\%$ of its initial mass, entirely supplied by the second cloud. Third, Fig.~\ref{fig:phase2}(b) shows that the depletion of $M_{c_2}$ is appreciable only for the smallest reservoir, $M_{c_2}=10^5\,\Msun$, which loses approximately $80\%$ of its initial mass. By contrast, the more massive reservoirs remain nearly unchanged, since spinning the BH up to $\chi_{\rm sat}$ requires only a limited amount of mass and angular momentum, largely independent of the total cloud mass available.

Second, our results indicate that the asymptotic state of the evolution is largely insensitive to the available cloud mass, if it contains enough angular momentum to reach the saturation spin. Although the adopted values of $M_{c_2}$ span a factor of fifty, both the final BH mass and the final spin vary by less than 10\%. Larger clouds extend the absorption process, which facilitates additional mass transfer before saturation is reached. However, because the endpoint is governed by the condition $\GammaSR\simeq0$ (or equivalently $\Omega_H=\mu$), the final state depends primarily on the coupling parameter $\al$ rather than the capacity of the cloud reservoir itself. In this sense, $M_{c_2}$ regulates the duration of Phase~2 but not its final outcome, closely mirroring the role played by the initial spin $\chi_0$ during Phase~1.

\begin{figure*}[!ht]
  \centering
  \includegraphics[width=\textwidth]{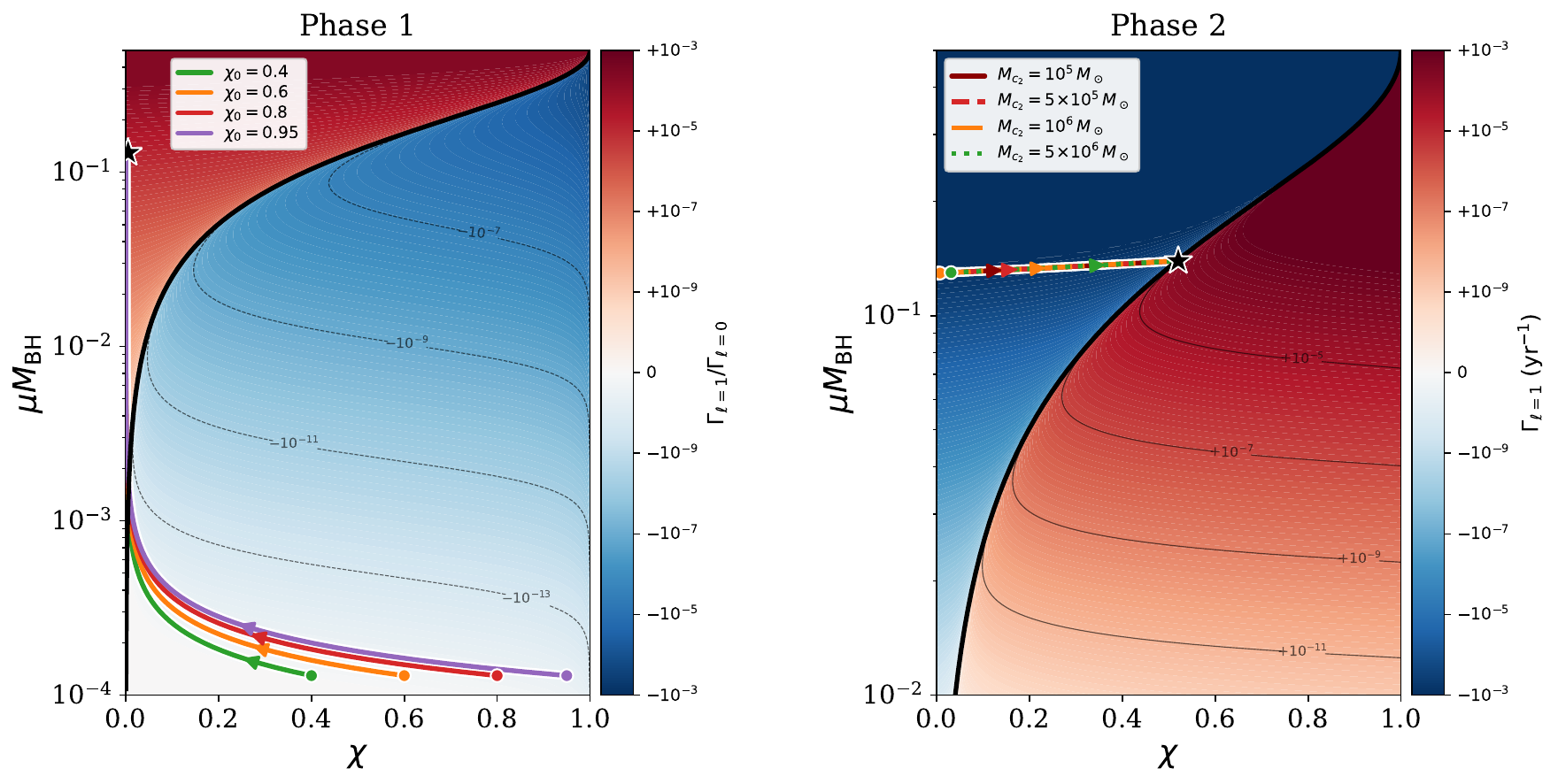}
  \caption{%
    Rate maps over the $(\chi,\,\mu\MBH)$ plane for
    $\mu = 1.73\times10^{-17}$~eV.  The solid black curve marks the saturation condition
$\chi_{\rm sat}^{(1)}(\mu\MBH)$ given by Eq.~\eqref{eq:chi_sat_general}, separating the absorption regime from the superradiant regime.
    \textbf{Left (Phase~1):} ratio $\GammaSR/\GammaA$ on a
    symmetric-log colour scale. Coloured solid curves show the Phase-1 trajectories for $\chi_0 \in \{0.4,0.6,0.8,0.95\}$ in the $(\chi,\mu\MBH)$ plane; arrows indicate the direction of evolution. All trajectories end in the upper-left region and converge to $(\chi\approx0,\,\mu\MBH\approx0.129)$ as the
    $\ell=0$ cloud is absorbed and the coupling grows. \textbf{Right (Phase~2):} signed $\GammaSR$ in yr$^{-1}$. Coloured curves show the Phase-2 trajectories for $M_{c_2}\in\{10^5,5\times10^5,10^6,5\times10^6\}\,\Msun$,
    all starting near $(\chi\approx0,\,\mu\MBH\approx0.129)$
    and sweeping rightward toward the superradiance boundary
    as the BH spin is driven to $\chi_{\rm final}\approx0.517$. The near-coincidence of the four curves demonstrates that
    the endpoint is set by $\mu\MBH$ rather than by $M_{c_2}$. Filled circles mark the initial points of each trajectory and stars mark the final equilibrium state reached at the saturation condition.}
  \label{fig3}
\end{figure*}

Third, as the BH approaches saturation, the accretion rate decreases by several orders of magnitude, while the coupling $\mu\MBH$ increases monotonically from $\simeq0.128$ to a constant value near $\simeq0.138$, as shown in Fig.~\ref{fig:phase2}(d). The vast majority of the mass transfer occurs before reaching the saturation threshold. In the limit $\GammaSR\rightarrow0$, both the accretion rate and the coupling become effectively constant, whereas the characteristic e-folding timescale diverges by several orders of magnitude. Consequently, the dynamical evolution naturally freezes once the system attains the saturation spin.

Finally, these findings reveal that the saturation mechanism is entirely self-regulating. Figure~\ref{fig:phase2}(f) shows that $\Gamma_{\rm SR}$ remains negative throughout the evolution and gradually approaches zero as the BH spin converges to $\chi_{\rm sat}$. For the more massive clouds, this convergence is accompanied by damped oscillations, before $\Gamma_{\rm SR}$ ultimately falls below the transition threshold $\Gamma_{\rm trans}$. No additional physical mechanism is required to halt the spin-up.
As in the standard superradiant instability, the evolution toward a
stationary hairy BH configuration is characterized by the
synchronization condition between the scalar field and the horizon:
the real part of the scalar-field frequency approaches
$\omega_R = m\Omega_H$, and the exchange of energy and angular
momentum between the BH and the scalar cloud is progressively
suppressed, vanishing at
synchronization~\cite{2017PhRvL.119d1101E}.

The key difference is that, whereas superradiance transfers energy and angular momentum from the BH to the scalar cloud, here the transfer proceeds from the scalar cloud to the BH.

Furthermore, we find that this equilibrium is dynamically stable. Any perturbation that increases the BH spin above $\chi_{\rm sat}$ renders $\Gamma_{l=1}>0$, thereby activating saturation spin extraction
and driving the system back toward equilibrium. A perturbation that decreases the spin below $\chi_{\rm sat}$ restores $\Gamma_{l=1}<0$, leading to renewed cloud absorption and spin-up.
These results identify $\chi_{\rm sat}$ as a genuine dynamical attractor, whose stability is demonstrated in more detail in Sec.~\ref{sec:phase2standalone}.

Having analyzed the two phases separately, we now examine the rate
hierarchy over the full $(\chi,\mu\MBH)$ parameter space to verify
the assumptions underlying the sequential treatment.
Figure~\ref{fig3} shows the corresponding results. The solid black curve denotes the saturation condition
$\chi_{\rm sat}^{(1)}(\mu\MBH)$ [Eq.~\eqref{eq:chi_sat_general}], separating the absorption and superradiant regimes of the $\ell=m=1$
mode.
As $\mu\MBH$ increases, the spin rises  monotonically, approaching $\chi\to1$ as $\mu\MBH\to1/2$.
 
The left panel confirms that the influence of superradiance during Phase~1 is negligible throughout the region traversed by the Phase~1 trajectories. Specifically, the ratio $|\GammaSR|/|\GammaA|$ remains extremely small, decreasing to $\lesssim10^{-13}$ at weak coupling. The overplotted Phase~1 trajectories (coloured curves, corresponding to the same initial spins as in Fig.~\ref{fig:phase1}) evolve from their initial spins toward $(\chi\approx0, \mu M_{\rm BH}\approx0.129)$ while remaining entirely within the absorption region where the $\ell=1$ superradiant mode is dynamically irrelevant. This confirms a posteriori that neglecting $\GammaSR$ in the Phase~1 evolution equations introduces a negligible correction, well below the per-mille level. This conclusion is not restricted to the fiducial parameters shown here: the same rate hierarchy persists throughout the displayed parameter space because $|\Gamma_{\ell=1}|/|\Gamma_{\ell=0}|\sim(\al)^5$ is parametrically suppressed in the small-coupling regime.
 
The right panel shows the signed superradiant rate, $\GammaSR$, whose zero defines the superradiance threshold. Its magnitude spans approximately twelve orders of magnitude across the displayed parameter space, reflecting the steep $(\mu M_{\rm BH})^9$ dependence of the rate. The overplotted Phase~2 trajectories (coloured curves, corresponding to the same initial cloud masses as in Fig.~\ref{fig:phase2}) all originate near $(\chi\approx0,,\mu M_{\rm BH}\approx0.129)$ in the absorption regime and evolve rightward toward the superradiance threshold as the BH spin increases to its saturation value, $\chi_{\rm final}\approx0.517$. Despite the initial cloud masses differing by a factor of fifty, the trajectories nearly overlap, demonstrating that the Phase~2 evolution is largely insensitive to the cloud reservoir. Consequently, the endpoint is determined primarily by the coupling $\mu M_{\rm BH}$ through the superradiance condition, rather than by the initial cloud mass, consistent with the time-domain evolution shown in Fig.~\ref{fig:phase2}.

\subsection{Deterministic spin imprinting of the BH}
\label{sec:imprinting}
The results of Secs.~\ref{sec:phase1} and~\ref{sec:phase2} can be summarized by a deterministic mapping from the initial BH spin to the final saturated state. Defining the \emph{imprinting map},
\begin{equation}
  \chi_0 \;\longmapsto\; \chi_{\rm final}
  \equiv \chi_{\rm sat}^{(1)}\!\left(\mu\MBH^{\rm end}\right),
  \label{eq:imprinting_map}
\end{equation}
where $\chi_{\rm sat}^{(1)}$ is given by Eq.~\eqref{eq:chi_sat_general} and $\MBH^{\rm end}$ denotes the BH mass at the end of Phase~2. We find that the final spin is independent of the initial spin $\chi_0$. Instead, $\chi_{\rm final}$ is determined entirely by the dimensionless coupling $\mu\MBH^{\rm end}$.

This result follows directly from the sequential evolution established in the previous sections. During Phase~1, the $\ell=0$ cloud erases the dependence on the initial BH spin by driving all configurations toward $\chi\simeq0$. During Phase~2, the subsequent absorption of the $\ell=m=1$ cloud spins the BH up to the saturation value $\chi_{\rm sat}^{(1)}(\mu\MBH^{\rm end})$, independently of its initial spin. The combined evolution therefore defines a spin-imprinting map that is largely independent of the BH's initial parameters.

The remaining dependence in Eq.~\eqref{eq:imprinting_map} is entirely contained in the coupling $\mu\MBH^{\rm end}$. In particular, the second-cloud mass $M_{c_2}$ is not an independent parameter of the map: its $\lesssim10\%$ effect on $\chi_{\rm final}$, shown in Fig.~\ref{fig:phase2}, arises only through the corresponding change in the final BH mass $\MBH^{\rm end}$. The final spin is therefore determined by the single combination $\mu\MBH$ evaluated at the end of the accretion process. For the fiducial parameters adopted in this work, $\mu=1.73\times10^{-17}\,\mathrm{eV}$ and $\MBH^{\rm end}\approx10^6\,\Msun$, we obtain $\chi_{\rm final}\approx0.52$.

More generally, Eq.~\eqref{eq:imprinting_map} defines a one-parameter family of predictions, assigning a unique final spin to each value of the coupling $\mu\MBH$. In Sec.~\ref{sec:saturation}, we evaluate this relation over the astrophysically relevant range of boson and BH masses, thereby constructing the saturation curve $\chi_{\rm sat}^{(1)}(\mu\MBH)$, which is subsequently compared with observed BH spins in Sec.~\ref{sec:obs}. The history independence established above is central to this comparison, as it implies that the observed final spin is determined solely by $\mu\MBH$, rather than by the unknown initial BH spin.

\section{Universal spin saturation and astrophysical implications}
\label{sec:saturation}
The two-phase evolution discussed in the previous section demonstrates that the final BH spin is not determined by its initial rotation, but by the superradiant equilibrium established during Phase~2. We now generalize this result beyond our fiducial example and derive the universal saturation spin for arbitrary BH masses, boson masses, and azimuthal modes. We then investigate the astrophysical conditions under which this equilibrium can be reached and whether it survives over cosmological times, we examine the role of higher-order modes and finally compare the predicted spin distribution with observations.

\begin{figure*}[t]
  \centering
  \includegraphics[width=1.8\columnwidth]{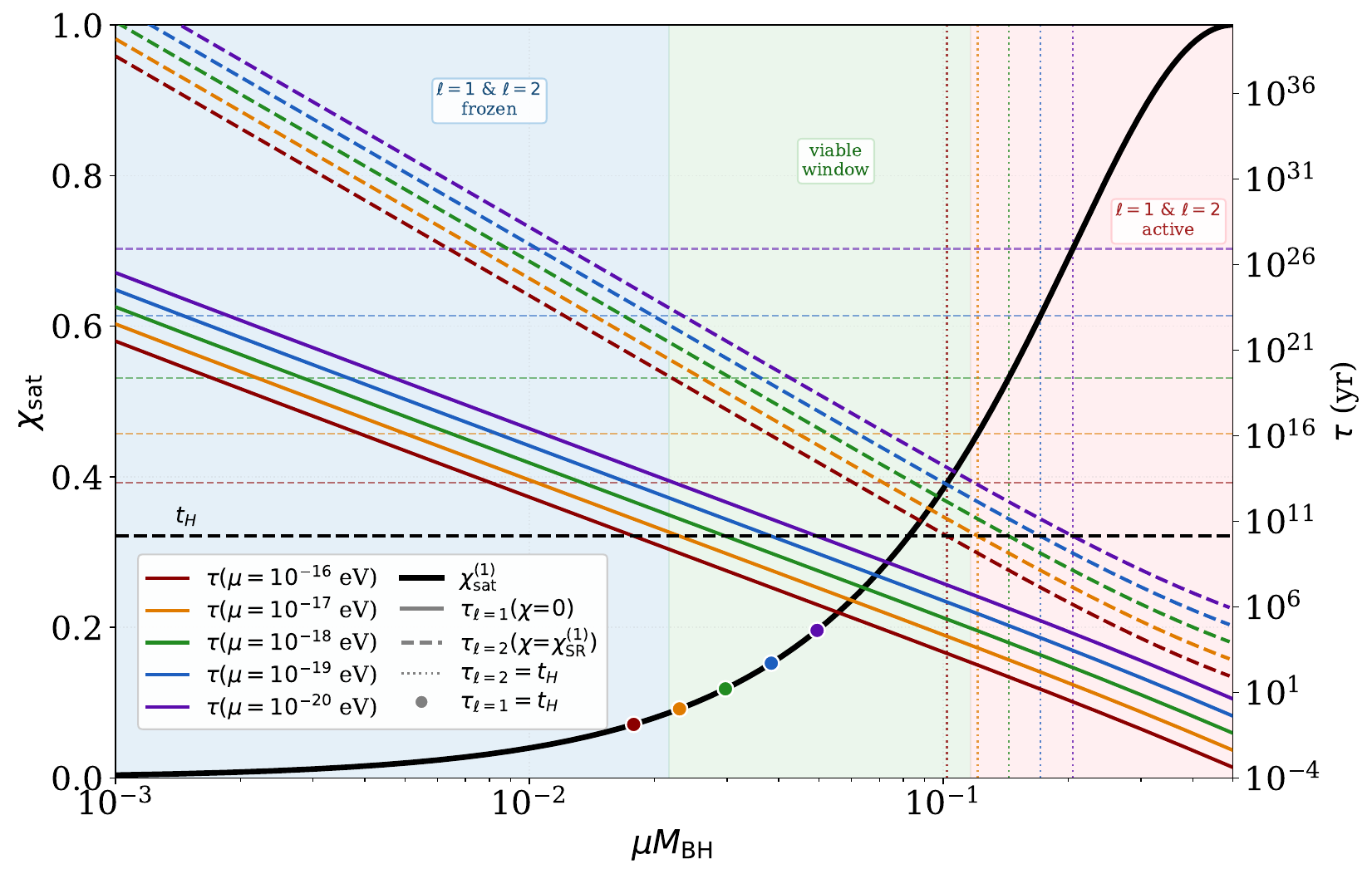}
  \caption{
    Unified diagram showing the saturation spin and characteristic timescales 
    for five selected boson masses $\mu \in [10^{-20},10^{-16}]$~eV.
    \textbf{Left axis, solid black:} $\ell=1$ saturation spin 
    $\chi_{\rm sat}^{(1)}(\al)$.
    \textbf{Right axis, solid colored:} growth timescale at the onset of 
    Phase 2, $\tau_{\ell=1}(\chi=0)$.
    \textbf{Right axis, dashed colored:} residual growth timescale for the 
    $\ell=2$ mode, $\tau_{\ell=2}(\chi_{\rm sat}^{(1)})$.
    \textit{Horizontal dashed black line:} age of the Universe, $t_H = 1.38\times10^{10}$~yr.
    \textit{Colored dots:} crossings where $\tau_{\ell=1}=t_H$, projected 
    onto the $\chi_{\rm sat}^{(1)}$ curve.
    \textit{Vertical dotted colored lines:} crossings where $\tau_{\ell=2}=t_H$.
    \textit{Horizontal colored dashes:} corresponding values of $\chi_{\rm sat}^{(1)}$ 
    at the $\tau_{\ell=2}=t_H$ transition for each mass $\mu$.
    The three shaded regions are defined with respect to the reference boson 
    mass $\mu_{\rm ref} = 1.73\times10^{-17}$~eV (see text):
    blue indicates the frozen regime ($\al < (\al)_{\rm crit}^{(1)} \approx 0.022$),
    green represents the viable window ($0.022 < \al < 0.116$),
    and red indicates the active $\ell=2$ regime ($\al > 0.116$).
    The quantitative parameters for each boson mass are listed in 
    Table~\ref{tab:crossings}.}
  \label{fig:merged_tau_chisat}
\end{figure*}

\begin{table*}[t]
\centering
\caption{
Critical couplings for which the characteristic interaction timescale equals
the age of the Universe,
$\tau_\ell=t_H=1.38\times10^{10}$ yr,
for representative boson masses.
The formation threshold corresponds to
$\tau_{\ell=1}(\chi=0)=t_H$,
whereas the stability threshold corresponds to
$\tau_{\ell=2}(\chi_{\rm sat}^{(1)})=t_H$.
The reference boson mass,
$\mu_{\rm ref}=1.73\times10^{-17}$ eV,
used to define the shaded regions in
Fig.~\ref{fig:merged_tau_chisat},
is marked with an asterisk.
}
\label{tab:crossings}
\begin{ruledtabular}
\begin{tabular}{lccccc}
Threshold &
$\mu$ (eV) &
$(\al)_{\rm crit}$ &
$M_{\rm crit}\,(M_\odot)$ &
$\chi_{\rm sat}^{(1)}$ &
$\chi_{\rm sat}^{(2)}$
\\
\colrule
Formation &
$10^{-16}$ & 0.0178 & $2.39\times10^{4}$ & 0.071 & 0.036\\
&
$1.73\times10^{-17 \star}$ & 0.0217 & $1.68\times10^{5}$ & 0.087 & 0.043\\
&
$10^{-17}$ & 0.0230 & $3.08\times10^{5}$ & 0.092 & 0.046\\
&
$10^{-18}$ & 0.0297 & $3.98\times10^{6}$ & 0.119 & 0.059\\
&
$10^{-19}$ & 0.0384 & $5.14\times10^{7}$ & 0.153 & 0.076\\
&
$10^{-20}$ & 0.0495 & $6.63\times10^{8}$ & 0.196 & 0.098\\
\colrule
Stability &
$10^{-16}$ & 0.1021 & $1.37\times10^{5}$ & 0.392 & 0.202\\
&
$1.73\times10^{-17 \star}$ & 0.1162 & $8.99\times10^{5}$ & 0.439 & 0.228\\
&
$10^{-17}$ & 0.1210 & $1.62\times10^{6}$ & 0.457 & 0.239\\
&
$10^{-18}$ & 0.1438 & $1.93\times10^{7}$ & 0.531 & 0.282\\
&
$10^{-19}$ & 0.1715 & $2.30\times10^{8}$ & 0.614 & 0.333\\
&
$10^{-20}$ & 0.2054 & $2.75\times10^{9}$ & 0.703 & 0.394\\
\end{tabular}
\end{ruledtabular}

\vspace{1mm}
{\footnotesize
$^{\star}$Reference boson mass adopted throughout this work and used to define
the shaded boundaries in Fig.~\ref{fig:merged_tau_chisat}.}
\end{table*}

\subsection{Universal saturation condition}
\label{sec:chi_sat}

In Sec.~\ref{sec:rates} we introduced the saturation condition $\GammaSR=0$, which defines $\chi_{\rm sat}^{(1)}$ via Eq.~\eqref{eq:chi_SR_def}. This condition generalizes immediately to a co-rotating mode of arbitrary azimuthal number $m$:
\begin{equation}
  \frac{\chi_{\rm sat}^{(m)}}{2\!\left(1+\sqrt{1-[\chi_{\rm sat}^{(m)}]^2}\right)}
  = \frac{\al}{m},
  \label{eq:chi_sat_m}
\end{equation}
which for $m=1$ reduces exactly to Eq.~\eqref{eq:chi_sat_general} and gives $\chi_{\rm sat}^{(1)}\simeq0.485$ for the fiducial values $\mu=1.73\times10^{-17}$~eV and $\MBH=10^6\,\Msun$ quoted in Sec.~\ref{sec:rates}. The left-hand side of Eq.~\eqref{eq:chi_sat_m} is $\Omega_H M_{\rm BH}$, a strictly increasing function of $\chi$ that vanishes at $\chi=0$ and approaches $1/2$ in the extremal limit $\chi\to1$. Within the approximation $\omega_R\simeq\mu$ adopted throughout this work, the right-hand side is simply the constant $\al$, so that a solution $\chi_{\rm sat}^{(m)}\in(0,1)$ exists if and only if $\al<1/2$. More generally, however, the exact synchronization condition is $\omega_R=m\Omega_H$, where the real part of the quasi-bound-state frequency satisfies $\omega_R\neq\mu$ once higher-order corrections are included. The limiting value $\al=m/2$ is itself approximate and is expected to receive corrections beyond the hydrogenic regime. This should not be interpreted as a limit on the existence of quasi-bound states, which persist beyond $\al=m/2$, but rather as a consequence of the approximation $\omega_R\simeq\mu$ used to derive Eq.~\eqref{eq:chi_sat_m}.

Equation~\eqref{eq:chi_sat_m} admits the closed-form solution
\begin{equation}
\chi_{\rm sat}^{(m)}
= \frac{4\al/m}{1+\left(2\al/m\right)^2}, \end{equation} 
derived in Appendix~\ref{app:chi_sat_closed_form}. This expression immediately yields the asymptotic limits $\chi_{\rm sat}^{(m)}\approx4\al/m$ for $\al\ll m/2$ and $\chi_{\rm sat}^{(m)}\rightarrow1$ as $\al\rightarrow m/2^{-}$. Within the approximation $\omega_R\simeq\mu$, Eq.~\eqref{eq:chi_sat_m} admits no saturation solution for
$\alpha\geq m/2$, since the corresponding boson frequency exceeds the maximum horizon frequency, $m,\Omega_H^{\rm max}=m/(2M_{\rm BH})$, attainable by a Kerr BH. This defines the approximate cutoff mass
\begin{equation}
M_{\rm cut}^{(m)}\equiv\frac{m}{2\mu},
\end{equation}
above which the corresponding mode cannot reach the synchronization condition within the present treatment and therefore no longer contributes to the BH spin evolution. Since $M_{\rm cut}^{(m)}\propto m$, higher-order modes remain relevant over progressively broader BH mass ranges. In particular, the $\ell=m=2$ mode extends the accessible mass window by a factor of two,
\begin{equation}
M_{\rm cut}^{(2)}=2M_{\rm cut}^{(1)}.
\end{equation}

\subsection{Cosmological survival and stability of the bosonic clouds}
\label{sec:survival}

Whether the proposed two-phase spin-imprinting mechanism can play a role in 
astrophysical scenarios depends on whether the relevant cloud-BH 
interaction timescales are shorter than the age of the Universe, $t_H \simeq 1.38\times10^{10}$~yr. We address this requirement by evaluating the timescales of two different modes, each computed at the appropriate phase of the evolution:
\begin{align}
  \tau_{\ell=1}(\al) &\equiv \frac{1}{|\GammaSR(\chi=0)|},
  \label{eq:tau1_survival}\\[4pt]
  \tau_{\ell=2}(\al) &\equiv
    \frac{1}{\left|\Gamma_{\ell=2}\!\left(\chi_{\rm sat}^{(1)}\right)\right|}.
  \label{eq:tau2_survival}
\end{align}
The timescale $\tau_{\ell=1}$ represents the characteristic decay (e-folding) time of the $\ell=m=1$ cloud at the beginning of Phase~2, when the BH spin is negligible ($\chi\simeq0$). It determines whether the BH can be spun up to the saturation value $\chi_{\rm sat}^{(1)}$ within a Hubble time. By contrast, $\tau_{\ell=2}$ is the characteristic superradiant growth (e-folding) time of a $\ell=m=2$ perturbation evaluated at the end of Phase~2, assuming that the $\ell=1$ mode has saturated ($\chi=\chi_{\rm sat}^{(1)}$ and $\GammaSR=0$). This timescale determines whether the $\ell=m=2$ mode can grow efficiently enough to destabilize the synchronized BH-cloud configuration by extracting angular momentum from the BH after Phase~2 has concluded~\cite{Degollado:2018ypf}. Both timescales are fully determined by Eqs.~\eqref{eq:Gamma_l1}-\eqref{eq:Gamma_l2}, without introducing additional free parameters.

To illustrate the interplay between these timescales and the 
saturation spin $\chi_{\rm sat}^{(1)}$, we construct the unified diagram shown 
in Fig.~\ref{fig:merged_tau_chisat}. This plot shows three separate physical 
quantities on a common axis representing the gravitational coupling $\mu\MBH$: 
the universal saturation spin $\chi_{\rm sat}^{(1)}(\al)$ (solid black curve, 
left axis), the Phase-2 onset timescale $\tau_{\ell=1}(\chi=0)$ (solid colored 
curves, right axis), and the residual $\ell=2$ timescale $\tau_{\ell=2}(\chi_{\rm sat}^{(1)})$ 
(dashed colored curves, right axis), for five representative boson masses in the 
range $\mu \in [10^{-20},10^{-16}]$~eV. The horizontal dashed black line 
indicates the Hubble time $t_H$.

As illustrated in Fig.~\ref{fig:merged_tau_chisat}, the parameter space can  be categorized into three distinct dynamical regimes. The boundaries of the  shaded regions are constructed using the reference boson mass $\mu_{\rm ref} = 1.73\times10^{-17}$~eV.  This choice corresponds to the baseline parameter used throughout  Secs.~\ref{sec:twophase}-\ref{sec:phase2standalone}, where the representative  BH mass of $\MBH = 10^6\,\Msun$ has been examined in detail. This  approach ensures that the dynamical regimes illustrated in the diagram align  directly with the numerical and analytical results discussed in previous sections.  We note that each individual boson mass possesses its own unique set of transition  points (represented by the colored dots and vertical dotted lines); consequently,  the actual boundary of the viable window for any specific mass $\mu$ is  determined by its respective crossing points rather than those of the reference mass.

\begin{figure*}[!ht]
  \centering
  \includegraphics[width=0.8\textwidth]{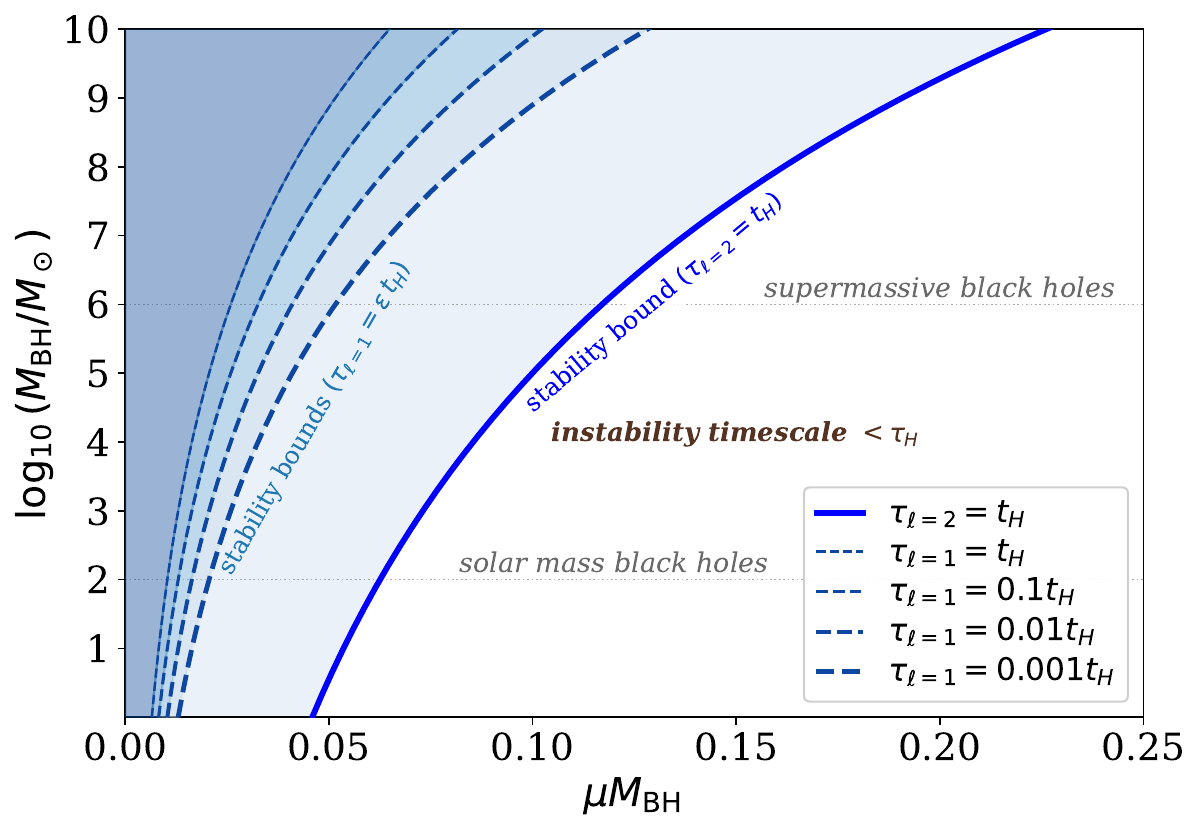}
  \caption{%
    Astrophysically viable domain in the
    $(\al,\,\log_{10}(\MBH/\Msun))$ plane.
    Solid curve: stability bound $\tau_{\ell=2}(\chi_{\rm sat}^{(1)})=t_H$;
    blue shading to the left indicates the $\ell=2$-frozen region.
    Dotted curves: formation bounds
    $\tau_{\ell=1}(\chi=0) = \varepsilon\,t_H$ for
    $\varepsilon = 1.0,\,0.1,\,0.01,\,0.001$.
    Darker shading between the stability bound $\tau_{\ell=2}$ and the
    $\varepsilon$ curves: the astrophysically viable domain
    where Phase~2 completes rapidly and the $\ell=2$ residue is
    cosmologically frozen.
  }
  \label{fig:stability_map}
\end{figure*}

\textit{Frozen regime}: For the reference mass, this boundary corresponds to a critical coupling of  $(\al)_{\rm crit}^{(1)} \approx 0.022$, or equivalently $\MBH \lesssim 1.7\times10^5\,\Msun$.  Below this coupling, the growth rate of the $\ell=1$ mode is sufficiently slow  that Phase~2 cannot be completed within a Hubble time, meaning the BH spin  retains a value close to its post-Phase-1 state over cosmological timescales. The colored  dots along the universal saturation curve indicate the shift in this threshold  for other boson masses: more massive bosons (higher $\mu$) become active at smaller BH masses (and hence smaller critical couplings $(\al)_{\rm crit}^{(1)}$) because the growth rate scales as $\GammaSR \propto \mu$, enhancing the interaction strength at a fixed coupling; the converse applies to lighter bosons.

\begin{figure*}[!ht]
  \centering
  \includegraphics[width=\textwidth]{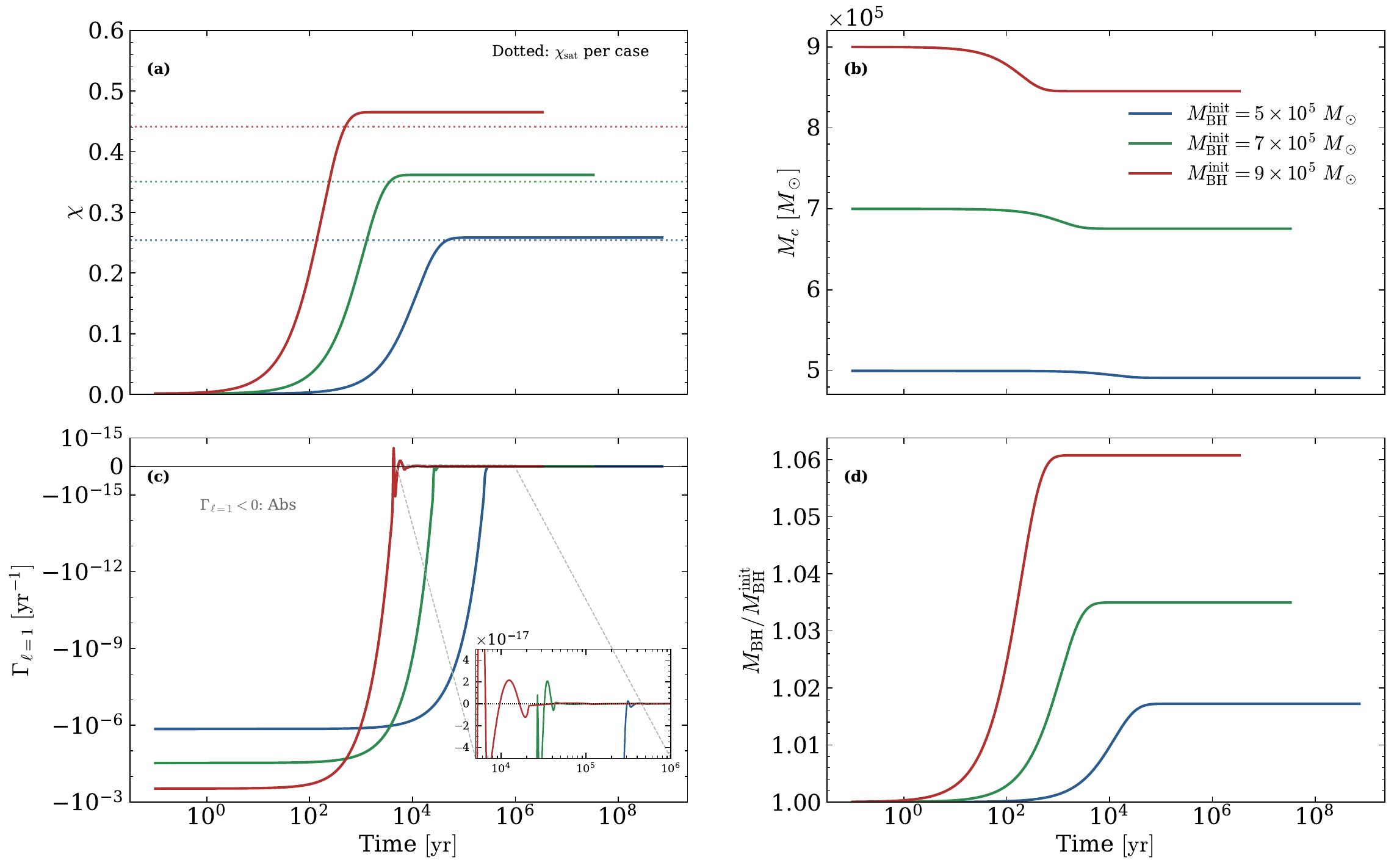}
  \caption{%
    Standalone Phase~2: a Schwarzschild BH ($\chi_0=0$) with
    $M_c = \MBH$ and $\mu = 1.73\times10^{-17}$~eV, for three initial
    masses $\MBH^{\rm init} \in \{5\times10^5,\,7\times10^5,\,9\times10^5\}
    \,\Msun$.
    \textbf{(a)} Spin $\chi(t)$; dotted horizontals mark
    $\chi_{\rm sat}\leq \chi_{\rm final}$ for each case.
    \textbf{(b)} Cloud mass $M_c(t)$.
    \textbf{(c)} Signed rate $\GammaSR(t)$; inset zooms on the
    damped oscillation about $\GammaSR=0$ at late times.
    \textbf{(d)} Fractional BH mass growth
    $\MBH/\MBH^{\rm init}$.
  }
  \label{fig:M2_standalone}
\end{figure*}

\textit{Viable window}:
For the reference mass, this window corresponds to the coupling range 
$0.022 \lesssim \al \lesssim 0.116$ (or $1.7\times10^5 \lesssim \MBH/\Msun \lesssim 9.0\times10^5$). 
Within this regime, Phase~2 is completed ($\tau_{\ell=1} < t_H$) while the 
$\ell=2$ perturbation remains cosmologically suppressed ($\tau_{\ell=2} > t_H$). 
Consequently, the spin prediction $\chi \approx \chi_{\rm sat}^{(1)}$ is 
both dynamically realizable and stable against higher-order mode corrections 
over the age of the Universe. Consistent with our previous choice, the upper 
boundary of this window is defined using $\mu_{\rm ref}$, as the long-term stability 
of the system against the $\ell=2$ residue has been explicitly confirmed for 
this mass in Sec.~\ref{sec:phase2}. The ratio between these two critical 
couplings, $(\al)_{\rm crit}^{(2)}/(\al)_{\rm crit}^{(1)} \approx 5.4$ for $\mu_{\rm ref}$, 
reflects the underlying timescale scaling $\tau_{\ell=2}/\tau_{\ell=1} \sim (\al)^{-5}$ 
given in Eq.~\eqref{eq:rate_ratio}. This implies that the $\ell=1$ mode becomes 
active at couplings roughly five orders of magnitude smaller than those required 
to activate the corresponding $\ell=2$ mode.

\textit{Active regime}:
For the reference mass, this regime is initiated at $(\al)_{\rm crit}^{(2)} \approx 0.116$ 
($\MBH \gtrsim 9.0\times10^5\,\Msun$). In this range, the $\ell=2$ mode can 
evolve within a Hubble time, introducing a perturbation to the saturation 
spin of order $\delta\chi/\chi_{\rm sat}^{(1)} \sim (\al)^5 \lesssim 10^{-2}$. 
Although this correction is parametrically small, it represents a potentially 
detectable feature for future high-precision spin measurements. This parameter 
space corresponds to the upper end of the BH mass range accessible via 
X-ray reflection spectroscopy in active galactic nuclei (AGN). The horizontal 
dashed colored lines show that the saturation spins at these crossings range 
from $\chi_{\rm sat}^{(1)} \approx 0.39$ for $\mu=10^{-16}$~eV to $\approx 0.70$ 
for $\mu=10^{-20}$~eV, identifying a specific target region for future observational tests. 
The critical couplings and corresponding physical parameters across all five 
investigated boson masses are summarized in Table~\ref{tab:crossings}.

\begin{figure*}[!ht]
  \centering
  \includegraphics[width=\textwidth]{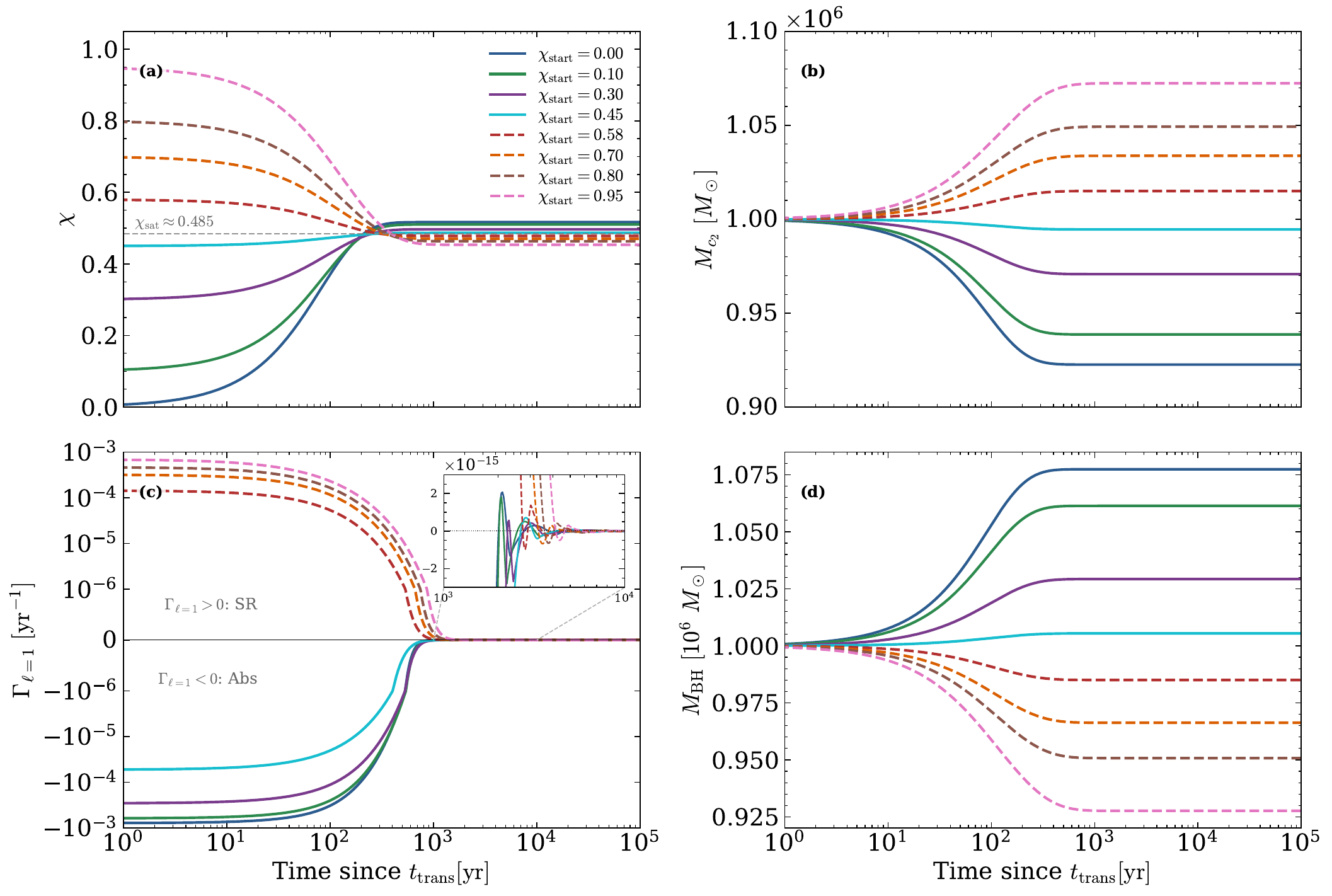}
\caption{%
    Phase~2 for $\mu = 1.73\times10^{-17}$~eV,
    $M_{\rm BH}^{\rm init} = 10^6\,\Msun$, $M_{c_2} = 10^6\,\Msun$,
    starting from eight initial spins
    $\chi_{\rm start} \in \{0.00,\,0.10,\,0.30,\,0.45,\,
    0.58,\,0.70,\,0.80,\,0.95\}$.
    Solid (cool-toned) curves: $\chi_{\rm start} < \chi_{\rm sat}$
    (absorption regime, spin driven up).
    Dashed (warm-toned) curves: $\chi_{\rm start} > \chi_{\rm sat}$
    (superradiant regime, spin driven down).
    \textbf{(a)} Spin $\chi(t)$; dashed horizontal marks
    $\chi_{\rm sat}$.
    \textbf{(b)} Cloud mass $M_{c_2}(t)$.
    \textbf{(c)} Signed rate $\Gamma_{\ell=1}(t)$; inset zooms on the
    damped oscillation about $\Gamma_{\ell=1}=0$ as $\chi\to\chi_{\rm sat}$.
    \textbf{(d)} BH mass in units of $10^6\,\Msun$.
  }
  \label{fig:B2_attractor}
\end{figure*}

Additionally, both critical couplings are displayed together in Fig.~\ref{fig:stability_map}. The resulting diagram is similar in spirit to the effective-stability map introduced in Ref.~\cite{Degollado:2018ypf}. The solid curve denotes the contour $\tau_{\ell=2}(\chi_{\rm sat}^{(1)})=t_H$, which separates the region where the $\ell=2$ mode remains dynamically suppressed from that where it can grow within a Hubble time. The dotted curves correspond to the formation contours $\tau_{\ell=1}(\chi=0)=\varepsilon,t_H$ for $\varepsilon\in\lbrace1,0.1,0.01,0.001\rbrace$. Throughout the mass range considered, these contours lie systematically to the left of the stability boundary, reflecting the hierarchy $|\GammaSR|\gg|\Gamma_{\ell=2}|$.

The region enclosed by the two families of curves identifies the parameter space in which the two-phase scenario operates as assumed throughout this work: the $\ell=m=1$ cloud spins the BH up to the synchronized configuration within a fraction $\varepsilon$ of the Hubble time, while the $\ell=m=2$ mode remains effectively inactive over cosmological timescales. For the reference case ($\MBH=10^6\,\Msun$ and $\mu=1.73\times10^{-17}$~eV), the coupling is $\mu\MBH\approx0.129$, which satisfies the formation bound with $\varepsilon \approx 9\times10^{-8}$ ($\tau_{\ell=1}\approx1.3\times10^3$~yr $\ll t_H$), confirming that Phase~2 completes almost instantaneously on cosmological timescales. However, $\al\approx0.129$ lies slightly beyond the stability boundary $\mu M_{\rm BH, stab}\approx0.116$ at this mass, so that the $\ell=m=2$ mode has a lifetime $\tau_{\ell=2}\approx3.3\times10^9$~yr $\approx0.24\,t_H$ and is therefore cosmologically active. The location of these boundaries follows directly from the analytical expressions in Eqs.~\eqref{eq:Gamma_l1}-\eqref{eq:Gamma_l2}, with $t_H$ as the only external input.

\subsection{Phase~2 as a standalone problem}
\label{sec:phase2standalone}

Having established the leading- and next-to-leading-order saturation
spins, we now investigate the Phase~2 evolution in isolation before
comparing the theoretical predictions with observational data. To this
end, we consider a Schwarzschild BH ($\chi_0=0$) with varying
initial mass interacting exclusively with an $\ell=m=1$ bosonic cloud.
This simplified configuration removes any dependence on the preceding
Phase~1 evolution, allowing the influence of the gravitational coupling
parameter $\al$ on both the saturation spin $\chi_{\rm sat}$ and the
characteristic spin-up timescale to be examined directly.

Figure~\ref{fig:M2_standalone} shows the resulting evolution for three
representative initial BH masses,
$M_{\rm BH}^{\rm init}\in
\{5\times10^5,\,
7\times10^5,\,
9\times10^5\}\,\Msun$,
assuming a boson mass
$\mu=1.73\times10^{-17}$~eV and an initial cloud mass
$M_c=M_{\rm BH}^{\rm init}$.
The spin-up timescale depends sensitively on the initial BH
mass. Since the analytical growth rate satisfies
$\Gamma_{\ell=1}\propto(\mu M_{\rm BH})^9$,
increasing the mass by a factor of $1.8$, from
$5\times10^5\,\Msun$ to $9\times10^5\,\Msun$, accelerates the evolution
by approximately $1.8^9\simeq200$, reducing the spin-up time from
$\sim10^6$~yr to $\sim10^3$~yr. The saturation spin likewise increases
monotonically with $\al$, reaching
$\chi_{\rm sat}\approx0.26$,
$0.36$, and
$0.47$ for
$M_{\rm BH}^{\rm init}=5\times10^5$,
$7\times10^5$, and
$9\times10^5\,\Msun$, respectively. Of these three cases, only the first two ($M_{\rm BH}^{\rm init}=5\times10^5$
and $7\times10^5\,\Msun$, with $\chi_{\rm sat}\approx0.26$ and $0.36$)
lie clearly below the $\ell=m=2$ instability threshold identified in
Sec.~\ref{sec:saturation}
($\chi_{\rm sat}^{(1)}\approx0.44$ at $\mu_{\rm ref}M_{\rm BH}\approx0.116$).
The heaviest case, $M_{\rm BH}^{\rm init}=9\times10^5\,\Msun$
($\chi_{\rm sat}\approx0.47$), lies essentially at this threshold, placing it at the boundary
between the cosmologically stable and $\ell=2$-active regimes.

\begin{figure*}[!ht]
  \centering
  \includegraphics[width=0.8\textwidth]{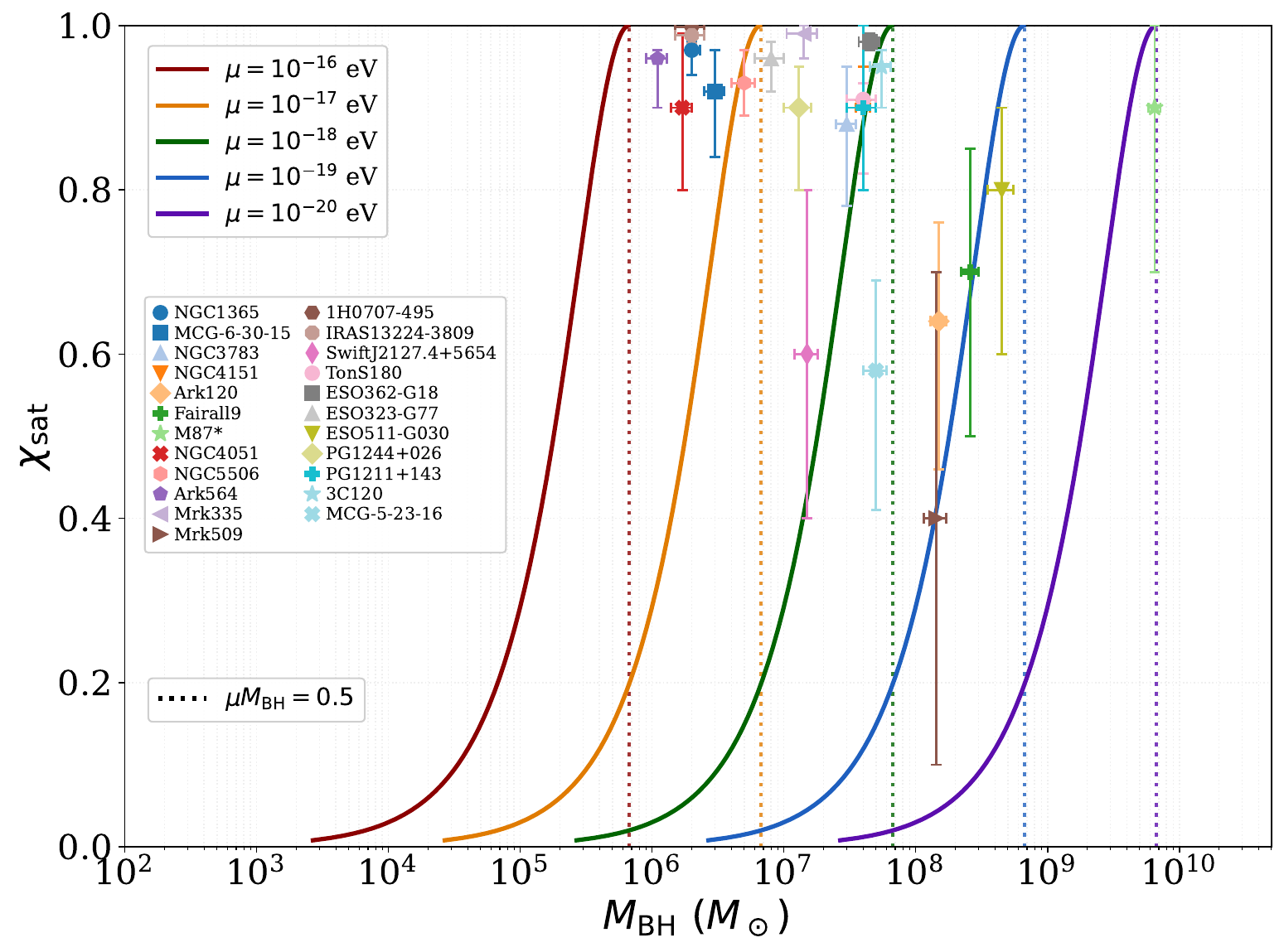}
  \caption{
    Superradiant saturation spin as a function of the BH mass
$\MBH$ for five representative boson masses,
$\mu\in[10^{-20},\,10^{-16}]~{\rm eV}$,
considering the $\ell=m=1$ mode in the regime $\al<0.5$.
The solid curves show the analytical prediction of
Eq.~\eqref{eq:chi_sat_general}, while the dotted vertical lines mark the
corresponding cutoff masses,
$M_{\rm cut}^{(1)}(\mu)$.
The overlaid data points denote AGN spin measurements inferred from
X-ray reflection spectroscopy
\cite{2014SSRv..183..277R,2021NatAs...5..133R}.
  }
  \label{fig:chi_sat_l12}
\end{figure*}

A notable feature is that each trajectory saturates slightly above its corresponding fixed-$\al$ equilibrium value $\chi_{\rm sat}$ (dotted lines). This behavior reflects the continuous growth of the BH mass through absorption (panel~(d)), which increases $\mu M_{\rm BH}$ during the evolution and consequently shifts the instantaneous equilibrium spin $\chi_{\rm sat}(\al)$ toward larger values. Although panel~(c) remains entirely within the absorptive regime ($\Gamma_{\ell=1}<0$), the inset shows that the approach to equilibrium is not strictly monotonic. Instead, $\Gamma_{\ell=1}$ undergoes a brief, weakly damped oscillation with an amplitude of order $10^{-17}\,{\rm yr}^{-1}$, including small positive (transiently superradiant) excursions, before ultimately relaxing to zero. However, note that this feature is sensitive to the integrator tolerance and disappears once the absolute tolerance is set component-wise rather than uniformly across state variables of vastly different magnitude ($M_{\rm BH}, M_c \sim 10^6$ versus $\chi \sim \mathcal{O}(1)$), indicating that it originates from floating-point roundoff in $\chi$ near $\Gamma_{\ell=1}=0$ rather than from genuine physical overshoot. The convergence to $\chi_{\rm sat}$ should therefore be understood as monotonic at the level of precision relevant to our conclusions. 

The attractor nature of the saturation spin is illustrated more clearly in Fig.~\ref{fig:B2_attractor}, which presents the Phase~2 evolution for eight initial spin values spanning $\chi_{\rm start}\in[0.00,\,0.95]$ at a fixed initial BH mass $M_{\rm BH}^{\rm init}=10^6\,\Msun$. Configurations with $\chi_{\rm start}<\chi_{\rm sat}\approx0.485$ (solid curves: $0.00$, $0.10$, $0.30$, and $0.45$) lie in the absorptive regime ($\Gamma_{\ell=1}<0$) and spin up toward $\chi_{\rm sat}$, whereas those with $\chi_{\rm start}>\chi_{\rm sat}$ (dashed curves: $0.58$, $0.70$, $0.80$, and $0.95$) occupy the superradiant regime ($\Gamma_{\ell=1}>0$) and spin down toward the same equilibrium. Despite their widely different initial conditions, all eight trajectories converge onto an essentially identical evolution at nearly the same time, largely independent of $\chi_{\rm start}$, demonstrating that $\chi_{\rm sat}$ can act as a robust dynamical attractor.

For all initial spins, the magnitude of the interaction rate, $|\Gamma_{\ell=1}|$, decreases as $\chi\rightarrow\chi_{\rm sat}$. The convergence, however, is not strictly monotonic. The inset of panel~(c) shows that $\Gamma_{\ell=1}$ undergoes a transient, weakly damped oscillation with alternating sign and a characteristic amplitude of $\sim2\times10^{-15}\,{\rm yr}^{-1}$ before asymptotically approaching zero. Accordingly, the BH spin undergoes a small overshoot and rings about $\chi_{\rm sat}$ before settling into the equilibrium state, rather than approaching it asymptotically from a single side. All eight trajectories merge onto essentially the same curve by $t\sim3\times10^2$-$2\times10^3$~yr, confirming that $\chi_{\rm sat}$ acts as a robust dynamical attractor independent of the direction of approach.

The associated mass exchange, shown in panels~(b) and~(d), scales systematically with the initial displacement from equilibrium, $|\chi_{\rm start}-\chi_{\rm sat}|$. Trajectories initialized closest to the saturation spin ($\chi_{\rm start}=0.45$ and $0.58$) experience negligible mass transfer, whereas the largest changes occur for the most extreme initial spins, $\chi_{\rm start}=0.00$ and $0.95$. In these cases, $M_{\rm BH}$ and $M_{c_2}$ vary by as much as $\sim7$-$8\%$ in opposite directions, with the BH gaining (losing) mass as the cloud loses (gains) mass during spin-up (spin-down) evolution. Nevertheless, the accompanying mass redistribution remains subdominant to the much more rapid transfer of angular momentum, which drives all trajectories toward the common equilibrium spin $\chi_{\rm sat}$.

\subsection{Comparison with AGN spin measurements}
\label{sec:obs}
Having established the saturation spin and its attractor behavior in the preceding section, we now compare the theoretical predictions with observational data. Figure~\ref{fig:chi_sat_l12} shows the saturation-spin curves for five representative boson masses, $\mu\in\left[10^{-20},10^{-16}\right]$~eV, together with spin measurements of 23 AGNs inferred from X-ray reflection spectroscopy~\cite{2014SSRv..183..277R,2021NatAs...5..133R,2021SSRv..217...65B}. For each boson mass, the corresponding $\ell=m=1$ saturation curve, $\chi_{\rm sat}(M_{\rm BH})$, is displayed together with the cutoff mass $M_{\rm cut}^{(1)}(\mu)$, defined within the approximation $\omega_R\simeq\mu$ by $\mu M_{\rm BH}=0.5$ and indicated by the dotted vertical line of the corresponding color.

Since $\chi_{\rm sat}$ depends on $M_{\rm BH}$ only through the dimensionless combination $\mu M_{\rm BH}$, the five curves do not represent independent predictions. Instead, varying $\mu$ continuously fills a region in the $(M_{\rm BH},\chi)$ plane, bounded above by the limiting behavior $\chi_{\rm sat}\rightarrow1$ as $\mu M_{\rm BH}\rightarrow1/2$ within the approximation adopted here. The mass-spin measurements of the 23 AGNs lie within this region, indicating that each source is consistent with the model for an appropriate value of the boson mass in the range considered. The observed spins span approximately $\chi\in[0.4,1.0]$, corresponding to the portion of the saturation curves with $\mu M_{\rm BH}\sim0.05$-$0.45$.

Several objects with near-extremal spins, including 1H0707-495, IRAS13224-3809, and Ark564, lie close to the high-$\mu M_{\rm BH}$ end of the $\ell=m=1$ saturation curves, where $\chi_{\rm sat}$ approaches unity. By contrast, the most massive objects in the sample, M87$^*$, ESO511-G030, and Fairall~9, lie beyond the $\ell=m=1$ cutoff for all but the smallest boson masses considered. Their observed spins may nevertheless be compatible with the broader BH mass range over which the $\ell=m=2$ mode can satisfy the synchronization condition, corresponding to $\mu\sim10^{-19}$-$10^{-18}$~eV.

The self-similar structure of the saturation curves also allows the boson mass to be inferred from a BH mass and spin measurement. Since $\chi_{\rm sat}^{(m)}$ depends only on the dimensionless coupling $\mu M_{\rm BH}$, a measurement of these quantities determines
\begin{equation}
\mu=m\Omega_H(\chi_{\rm sat}^{(m)}),
\label{eq:mu_constraint}
\end{equation}
through the saturation condition, Eq.~\eqref{eq:chi_sat_m}, where $\Omega_H$ is evaluated from the measured BH mass and spin using Eq.~\eqref{eq:OmH}. Within the framework of the present model, this inference is independent of the BH formation history. Since changing $\mu$ simply rescales the saturation curves along the BH mass axis while preserving their functional form, measurements of BH masses and spins can, in principle, be used to constrain the boson mass. More stringent constraints will require improved BH mass and spin measurements, together with more accurate X-ray reflection spectroscopy
models used to fit the relativistically broadened iron line
profiles from which BH spins are
inferred~\cite{2014SSRv..183..277R,2021NatAs...5..133R}. The comparison indicates that several observed mass-spin pairs are compatible with the predicted saturation curves for suitable values of $\mu$, while others lie outside the parameter space covered by the model. Due to the simplified nature of the present framework, this comparison is only a preliminary consistency check and cannot be regarded as observational evidence for the proposed mechanism. The present analysis can be further extended by incorporating baryonic accretion, BH mergers, environmental effects, gravitational backreaction, and the possible $\ell=m=2$ instability discussed in Sec.~\ref{sec:saturation}. These effects may modify the predicted mass-spin relation and provide a more comprehensive assessment of the model.

\section{Conclusions}
\label{sec:conclusions}
In this work, we have developed an analytical model for the coupled evolution of a Kerr BH interacting with two ultralight scalar clouds occupying different quasi-bound states. By combining the absorption and superradiant rates with a self-consistent evolution of the BH mass and spin, we followed the system through two successive phases: an initial stage of rapid mass growth driven by the $\ell=0$ cloud, followed by the accretion of the $\ell=m=1$ cloud, which spins the BH up to the synchronization threshold. This sequential evolution naturally links the BH mass growth and spin evolution within a single framework. Our results show that the final BH spin is largely independent of the initial spin and is instead determined primarily by the synchronization condition, providing a simple spin-imprinting mechanism controlled by the boson mass and the final BH mass.

The first evolutionary stage describes the absorption of the spherically symmetric ($\ell=0$) scalar cloud. Due to the strong dependence of the absorption rate on the BH mass, the evolution develops into a runaway growth phase that allows BH seeds of order $10^3\,M_\odot$ to reach supermassive scales on cosmologically relevant timescales. At the same time, the accretion of the non-rotating cloud reduces the dimensionless BH spin, driving the system toward $\chi\simeq0$ largely independently of its initial value. This evolution erases the memory of the initial spin and provides the initial conditions for the subsequent phase.

During the second stage, the absorption of the $\ell=m=1$ cloud transfers both mass and angular momentum to the BH, increasing its spin until the saturation condition is reached. As the BH approaches this state, the absorption rate decreases continuously and vanishes at the superradiant threshold, leading the system to a stable equilibrium without requiring an additional saturation mechanism. The resulting saturation spin is determined primarily by the dimensionless coupling $\mu M_{\rm BH}$ and depends only weakly on the previous evolution. Consequently, the model predicts a family of saturation curves in the BH mass-spin plane, parameterized by the boson mass.

Finally, we extended the analysis to higher-$m$ superradiant modes. We derived the corresponding analytical rates and synchronization conditions for arbitrary azimuthal number $m$, together with the associated cutoff masses. As an application, we considered the $\ell=m=2$ mode and showed that it extends the BH mass range over which synchronization can occur while remaining strongly suppressed relative to the $\ell=m=1$ mode throughout most of the perturbative regime. The corresponding growth timescales indicate that the $\ell=m=2$ instability becomes relevant only near the upper end of the parameter space considered here. This mode is nevertheless of particular interest, as it represents the leading instability channel of the synchronized BH-cloud configuration and may drive its evolution away from the $\ell=m=1$ equilibrium~\cite{Degollado:2018ypf}.

Beyond describing the two-phase evolution, the present framework establishes a direct connection between the model and astrophysical observations. By combining the characteristic timescales associated with the completion of Phase~2 and the growth of the $\ell=m=2$ mode, we identified the region of parameter space in which the predicted saturation curves can be established and remain stable over cosmological timescales. This region follows directly from the analytical evolution equations without introducing additional phenomenological assumptions. Comparison with the current AGN sample shows that several measured BH mass-spin pairs may be compatible with the predicted saturation curves for suitable boson masses, while others fall outside the simplified evolutionary picture considered here. Given the idealized nature of the present model, this level of agreement should be regarded as encouraging rather than conclusive. Additional physical processes are expected to broaden the range of possible evolutionary tracks in the BH mass-spin plane and should be taken into account before quantitative constraints on the boson mass can be inferred.

It is important to remark again on the nature of the endpoint to avoid misconceptions regarding similar systems reported in the literature. At the end of Phase~2, the system consists of a rotating SMBH surrounded by an angular-momentum-carrying bosonic scalar cloud. The evolution drives the system toward the superradiant threshold, where the net scalar flux through the horizon vanishes and the accretion process saturates. Within our prescription, the resulting configuration is a long-lived, stationary-like black hole surrounded by a scalar cloud. However, this endpoint should not be identified with an exact black hole with synchronized scalar hair solution \cite{Herdeiro:2014goa}, since such configurations are fully backreacting, stationary, and axisymmetric solutions of the Einstein-Klein-Gordon equations. In a hairy Kerr black hole, the spacetime geometry and scalar field are determined self-consistently under the appropriate synchronization condition. By contrast, our model describes the evolution of a Kerr black hole interacting with quasi-bound scalar clouds through analytical absorption and superradiant rates. It therefore establishes the saturation of the evolutionary process but does not construct the corresponding nonlinear equilibrium solution. The resulting configuration is closer to the {\it wig} configuration of a Schwarzschild black hole \cite{barranco2011black} and may represent a perturbative-like precursor to, or an astrophysically motivated pathway toward, a black hole with scalar hair. Whether it evolves into such a solution, however, cannot be determined within the present framework.

Several extensions of the present framework remain to be explored. These include the incorporation of gravitational backreaction through the fully nonlinear Einstein-Klein-Gordon system~\cite{2018PhRvL.121m1104E,2017PhRvL.119d1101E}, as well as baryonic accretion, black hole mergers, environmental interactions, and repeated episodes of cloud formation. A more realistic treatment of the initial populations of quasi-bound clouds in a cosmological setting would further improve the physical modeling.

\begin{acknowledgments}
SF is funded by the Conselleria de Innovaci\'on, Universidades, Ciencia y Sociedad Digital of the Generalitat Valenciana and the European Social Fund through a postdoctoral fellowship APOSTD 2025 (CIAPOS/2024/461). NSG acknowledges support from the Spanish Ministry of Science and Innovation via the Ram\'on y Cajal programme (grant RYC2022-037424-I), funded by MCIN/AEI/10.13039/501100011033 and by ``ESF Investing in your future". JCM is funded by a UNAM-DGAPA Postdoctoral Fellowship and by the PAPIIT-IN108526 Determinaci\'on de consecuencias observacionales de la naturaleza
de las componentes obscuras del Universo project. This work is also supported by the Spanish Agencia Estatal de Investigación (grant PID2024-159689NB-C21) funded by MICIU/AEI/10.13039/501100011033 and by FEDER/EU, by the Generalitat Valenciana (Prometeo grant CIPROM/2022/49), by PAPIIT grant IN107026, and by the European Horizon Europe staff exchange (SE) programme HORIZON-MSCA2021-SE-01 Grant No. NewFunFiCO-101086251. SF and NSG gratefully acknowledge Rodrigo Vicente for valuable discussions and comments that contributed to this work.
\end{acknowledgments}

\bibliographystyle{apsrev4-2}
\bibliography{draft_ml}

@ARTICLE{2026PhLB..87440251S,
       author = {{Sanchis-Gual}, Nicolas and {Barranco}, Juan and {Degollado}, Juan Carlos and {N{\'u}{\~n}ez}, Dar{\'\i}o},
        title = "{Dark-to-black super accretion as a mechanism for early supermassive black hole growth}",
      journal = {Physics Letters B},
         year = 2026,
        month = mar,
       volume = {874},
          eid = {140251},
        pages = {140251},
          doi = {10.1016/j.physletb.2026.140251},
archivePrefix = {arXiv},
       eprint = {2510.00644},
 primaryClass = {astro-ph.CO},
       adsurl = {https://ui.adsabs.harvard.edu/abs/2026PhLB..87440251S}
}

@article{Matos:1998vk,
    author = "Matos, Tonatiuh and Guzman, Francisco Siddhartha",
    title = "{Scalar fields as dark matter in spiral galaxies}",
    eprint = "gr-qc/9810028",
    archivePrefix = "arXiv",
    reportNumber = "CINVESTAV-98-22",
    doi = "10.1088/0264-9381/17/1/102",
    journal = "Class. Quant. Grav.",
    volume = "17",
    pages = "L9--L16",
    year = "2000"
}

@article{Alcubierre:2025zus,
    author = "Alcubierre, Miguel and Barranco, Juan and Bernal, Argelia and Degollado, Juan Carlos and Diez-Tejedor, Alberto and Megevand, Miguel and Nunez, Dario and Sarbach, Olivier",
    title = "{Noble gravitational atoms: self-gravitating black hole scalar wigs with angular momentum number}",
    eprint = "2512.08095",
    archivePrefix = "arXiv",
    primaryClass = "gr-qc",
    doi = "10.1088/1361-6382/ae4160",
    journal = "Class. Quant. Grav.",
    volume = "43",
    number = "4",
    pages = "045010",
    year = "2026"
}

@ARTICLE{2006AJ....131.1203F,
       author = {{Fan}, Xiaohui and {Strauss}, Michael A. and {Richards}, Gordon T. and {Hennawi}, Joseph F. and {Becker}, Robert H. and {White}, Richard L. and {Diamond-Stanic}, Aleksandar M. and {Donley}, Jennifer L. and {Jiang}, Linhua and {Kim}, J. Serena and {Vestergaard}, Marianne and {Young}, Jason E. and {Gunn}, James E. and {Lupton}, Robert H. and {Knapp}, Gillian R. and {Schneider}, Donald P. and {Brandt}, W.~N. and {Bahcall}, Neta A. and {Barentine}, J.~C. and {Brinkmann}, J. and {Brewington}, Howard J. and {Fukugita}, Masataka and {Harvanek}, Michael and {Kleinman}, S.~J. and {Krzesinski}, Jurek and {Long}, Dan and {Neilsen}, Jr., Eric H. and {Nitta}, Atsuko and {Snedden}, Stephanie A. and {Voges}, Wolfgang},
        title = "{A Survey of z>5.7 Quasars in the Sloan Digital Sky Survey. IV. Discovery of Seven Additional Quasars}",
      journal = {\aj},
         year = 2006,
        month = mar,
       volume = {131},
       number = {3},
        pages = {1203-1209},
          doi = {10.1086/500296},
archivePrefix = {arXiv},
       eprint = {astro-ph/0512080},
 primaryClass = {astro-ph},
       adsurl = {https://ui.adsabs.harvard.edu/abs/2006AJ....131.1203F}
}

@ARTICLE{2011Natur.474..616M,
       author = {{Mortlock}, Daniel J. and {Warren}, Stephen J. and {Venemans}, Bram P. and {Patel}, Mitesh and {Hewett}, Paul C. and {McMahon}, Richard G. and {Simpson}, Chris and {Theuns}, Tom and {Gonz{\'a}les-Solares}, Eduardo A. and {Adamson}, Andy and {Dye}, Simon and {Hambly}, Nigel C. and {Hirst}, Paul and {Irwin}, Mike J. and {Kuipers}, Ernst and {Lewis}, Antonia and {Middleton}, Max},
        title = "{A luminous quasar at a redshift of z = 7.085}",
      journal = {\nat},
         year = 2011,
       volume = {474},
       number = {7353},
        pages = {616-619},
          doi = {10.1038/nature10159},
archivePrefix = {arXiv},
       eprint = {1106.6088},
       adsurl = {https://ui.adsabs.harvard.edu/abs/2011Natur.474..616M}
}

@ARTICLE{2015Natur.518..512W,
       author = {{Wu}, Xue-Bing and {Wang}, Feige and {Fan}, Xiaohui and {Yi}, Weimin and {Zuo}, Wenwen and {Bian}, Fuyan and {Jiang}, Linhua and {McGreer}, Ian D. and {Wang}, Ran and {Yang}, Jinyi and {Yang}, Qian and {Thompson}, David and {Beletsky}, Yuri},
        title = "{An ultraluminous quasar with a twelve-billion-solar-mass black hole at redshift 6.30}",
      journal = {\nat},
         year = 2015,
       volume = {518},
       number = {7540},
        pages = {512-515},
          doi = {10.1038/nature14241},
archivePrefix = {arXiv},
       eprint = {1502.07418},
       adsurl = {https://ui.adsabs.harvard.edu/abs/2015Natur.518..512W}
}

@ARTICLE{2018Natur.553..473B,
       author = {{Ba{\~n}ados}, Eduardo and {Venemans}, Bram P. and {Mazzucchelli}, Chiara and {Farina}, Emanuele P. and {Walter}, Fabian and {Wang}, Feige and {Decarli}, Roberto and {Stern}, Daniel and {Fan}, Xiaohui and {Davies}, Frederick B. and {Hennawi}, Joseph F. and {Simcoe}, Robert A. and {Turner}, Monica L. and {Rix}, Hans-Walter and {Yang}, Jinyi and {Kelson}, Daniel D. and {Rudie}, Gwen C. and {Winters}, Jan Martin},
        title = "{An 800-million-solar-mass black hole in a significantly neutral Universe at a redshift of 7.5}",
      journal = {\nat},
         year = 2018,
       volume = {553},
       number = {7689},
        pages = {473-476},
          doi = {10.1038/nature25180},
archivePrefix = {arXiv},
       eprint = {1712.01860},
       adsurl = {https://ui.adsabs.harvard.edu/abs/2018Natur.553..473B}
}

@ARTICLE{2021ApJ...907L...1W,
       author = {{Wang}, Feige and {Yang}, Jinyi and {Fan}, Xiaohui and {Hennawi}, Joseph F. and {Barth}, Aaron J. and {Banados}, Eduardo and {Bian}, Fuyan and {Boutsia}, Konstantina and {Connor}, Thomas and {Davies}, Frederick B. and {Decarli}, Roberto and {Eilers}, Anna-Christina and {Farina}, Emanuele P. and {Green}, Richard and {Jiang}, Linhua and {Li}, Jiang-Tao and {Mazzucchelli}, Chiara and {Pan}, Lile and {Schindler}, Jan-Torge and {Venemans}, Bram and {Walter}, Fabian and {Wu}, Xue-Bing and {Yue}, Minghao},
        title = "{A Luminous Quasar at Redshift 7.642}",
      journal = {\apjl},
         year = 2021,
       volume = {907},
       number = {1},
          eid = {L1},
        pages = {L1},
          doi = {10.3847/2041-8213/abd8c6},
archivePrefix = {arXiv},
       eprint = {2101.03179},
       adsurl = {https://ui.adsabs.harvard.edu/abs/2021ApJ...907L...1W}
}

@ARTICLE{2021ApJ...908L..33Y,
       author = {{Yang}, Jinyi and {Wang}, Feige and {Fan}, Xiaohui and {Hennawi}, Joseph F. and {Davies}, Frederick B. and {Yue}, Minghao and {Banados}, Eduardo and {Wu}, Xue-Bing and {Venemans}, Bram and {Barth}, Aaron J. and {Bian}, Fuyan and {Boutsia}, Konstantina and {Connor}, Thomas and {Decarli}, Roberto and {Eilers}, Anna-Christina and {Farina}, Emanuele P. and {Green}, Richard and {Jiang}, Linhua and {Li}, Jiang-Tao and {Mazzucchelli}, Chiara and {Pan}, Lile and {Schindler}, Jan-Torge and {Walter}, Fabian},
        title = "{Poniua'ena: A Luminous z = 7.515 Quasar Hosting a 1.5 Billion Solar Mass Black Hole}",
      journal = {\apjl},
         year = 2021,
       volume = {908},
       number = {1},
          eid = {L33},
        pages = {L33},
          doi = {10.3847/2041-8213/abe084},
archivePrefix = {arXiv},
       eprint = {2006.13452},
       adsurl = {https://ui.adsabs.harvard.edu/abs/2021ApJ...908L..33Y}
}

@ARTICLE{2010A&ARv..18..279V,
       author = {{Volonteri}, Marta},
        title = "{Formation of supermassive black holes}",
      journal = {\aapr},
         year = 2010,
       volume = {18},
       number = {3},
        pages = {279-315},
          doi = {10.1007/s00159-010-0029-x},
archivePrefix = {arXiv},
       eprint = {1003.4404},
       adsurl = {https://ui.adsabs.harvard.edu/abs/2010A%26ARv..18..279V}
}

@ARTICLE{2020ARA&A..58...27I,
       author = {{Inayoshi}, Kohei and {Visbal}, Eli and {Haiman}, Zolt{\'a}n},
        title = "{The Assembly of the First Massive Black Holes}",
      journal = {\araa},
         year = 2020,
       volume = {58},
        pages = {27-97},
          doi = {10.1146/annurev-astro-120419-014455},
archivePrefix = {arXiv},
       eprint = {1911.05791},
       adsurl = {https://ui.adsabs.harvard.edu/abs/2020ARA%26A..58...27I}
}

@ARTICLE{2001ApJ...551L..27M,
       author = {{Madau}, Piero and {Rees}, Martin J.},
        title = "{Massive Black Holes as Population III Remnants}",
      journal = {\apjl},
         year = 2001,
       volume = {551},
       number = {1},
        pages = {L27-L30},
          doi = {10.1086/319848},
archivePrefix = {arXiv},
       eprint = {astro-ph/0101223},
       adsurl = {https://ui.adsabs.harvard.edu/abs/2001ApJ...551L..27M}
}

@ARTICLE{2004ApJ...604..484M,
       author = {{Madau}, Piero and {Rees}, Martin J. and {Volonteri}, Marta and {Haardt}, Francesco and {Oh}, S. Peng},
        title = "{Early Reionization by Miniquasars}",
      journal = {\apj},
         year = 2004,
       volume = {604},
       number = {2},
        pages = {484-494},
          doi = {10.1086/381935},
archivePrefix = {arXiv},
       eprint = {astro-ph/0310223},
       adsurl = {https://ui.adsabs.harvard.edu/abs/2004ApJ...604..484M}
}

@ARTICLE{2011Sci...334..948T,
       author = {{Turk}, Matthew J. and {Abel}, Tom and {O'Shea}, Brian},
        title = "{The Formation of Population III Binaries from Cosmological Initial Conditions}",
      journal = {Science},
         year = 2009,
       volume = {325},
        pages = {601},
          doi = {10.1126/science.1173540},
archivePrefix = {arXiv},
       eprint = {0907.2919},
       adsurl = {https://ui.adsabs.harvard.edu/abs/2009Sci...325..601T}
}

@ARTICLE{1994ApJ...432...52L,
       author = {{Loeb}, Abraham and {Rasio}, Frederic A.},
        title = "{Collapse of Primordial Gas Clouds and the Formation of Quasar Black Holes}",
      journal = {\apj},
         year = 1994,
       volume = {432},
        pages = {52-61},
          doi = {10.1086/174548},
archivePrefix = {arXiv},
       eprint = {astro-ph/9401026},
       adsurl = {https://ui.adsabs.harvard.edu/abs/1994ApJ...432...52L}
}

@ARTICLE{2006MNRAS.370..289B,
       author = {{Begelman}, Mitchell C. and {Volonteri}, Marta and {Rees}, Martin J.},
        title = "{Massive black hole seeds born with super-Eddington accretion}",
      journal = {\mnras},
         year = 2006,
       volume = {370},
       number = {1},
        pages = {289-298},
          doi = {10.1111/j.1365-2966.2006.10467.x},
archivePrefix = {arXiv},
       eprint = {astro-ph/0602363},
       adsurl = {https://ui.adsabs.harvard.edu/abs/2006MNRAS.370..289B}
}

@ARTICLE{2010MNRAS.402.1249S,
       author = {{Shang}, Cien and {Bryan}, Greg L. and {Haiman}, Zolt{\'a}n},
        title = "{Supermassive black hole formation at high redshift triggered by coevolution of an ultraviolet radiation field and cold pristine gas}",
      journal = {\mnras},
         year = 2010,
       volume = {402},
       number = {2},
        pages = {1249-1262},
          doi = {10.1111/j.1365-2966.2009.15960.x},
archivePrefix = {arXiv},
       eprint = {0906.0267},
       adsurl = {https://ui.adsabs.harvard.edu/abs/2010MNRAS.402.1249S}
}

@ARTICLE{2017MNRAS.469.3329W,
       author = {{Wolcott-Green}, Jemma and {Haiman}, Zolt{\'a}n and {Bryan}, Greg L.},
        title = "{Suppression of H$_{2}$ cooling in protogalactic clouds by Lyman-Werner radiation: 3D effects}",
      journal = {\mnras},
         year = 2017,
       volume = {469},
       number = {3},
        pages = {3329-3341},
          doi = {10.1093/mnras/stx1107},
archivePrefix = {arXiv},
       eprint = {1611.09363},
       adsurl = {https://ui.adsabs.harvard.edu/abs/2017MNRAS.469.3329W}
}

@ARTICLE{2004Natur.428..724P,
       author = {{Portegies Zwart}, Simon F. and {Baumgardt}, Holger and {Hut}, Piet and {Makino}, Junichiro and {McMillan}, Stephen L. W.},
        title = "{Formation of massive black holes through runaway collisions in dense young star clusters}",
      journal = {\nat},
         year = 2004,
       volume = {428},
       number = {6984},
        pages = {724-726},
          doi = {10.1038/nature02448},
archivePrefix = {arXiv},
       eprint = {astro-ph/0402622},
       adsurl = {https://ui.adsabs.harvard.edu/abs/2004Natur.428..724P}
}

@ARTICLE{2010ARA&A..48..339B,
       author = {{Bromm}, Volker and {Yoshida}, Naoki},
        title = "{The First Galaxies}",
      journal = {\araa},
         year = 2011,
       volume = {49},
        pages = {373-407},
          doi = {10.1146/annurev-astro-120210-163243},
archivePrefix = {arXiv},
       eprint = {1102.4638},
       adsurl = {https://ui.adsabs.harvard.edu/abs/2011ARA%26A..49..373B}
}

@ARTICLE{2000PhRvL..85.1158H,
       author = {{Hu}, Wayne and {Sawicki}, Ignacy and {Frieman}, Joshua},
        title = "{Fuzzy Cold Dark Matter: The Wave Properties of Ultralight Particles}",
      journal = {\prl},
         year = 2000,
       volume = {85},
       number = {6},
        pages = {1158-1161},
          doi = {10.1103/PhysRevLett.85.1158},
archivePrefix = {arXiv},
       eprint = {astro-ph/0003365},
       adsurl = {https://ui.adsabs.harvard.edu/abs/2000PhRvL..85.1158H}
}

@ARTICLE{2017PhRvD..95d3541H,
       author = {{Hui}, Lam and {Ostriker}, Jeremiah P. and {Tremaine}, Scott and {Witten}, Edward},
        title = "{Ultralight axions in astronomy and cosmology}",
      journal = {\prd},
         year = 2017,
       volume = {95},
       number = {4},
          eid = {043541},
        pages = {043541},
          doi = {10.1103/PhysRevD.95.043541},
archivePrefix = {arXiv},
       eprint = {1610.08297},
       adsurl = {https://ui.adsabs.harvard.edu/abs/2017PhRvD..95d3541H}
}

@ARTICLE{2010PhRvD..81l3530A,
       author = {{Arvanitaki}, Asimina and {Dimopoulos}, Savas and {Dubovsky}, Sergei and {Kaloper}, Nemanja and {March-Russell}, John},
        title = "{String axiverse}",
      journal = {\prd},
         year = 2010,
       volume = {81},
       number = {12},
          eid = {123530},
        pages = {123530},
          doi = {10.1103/PhysRevD.81.123530},
archivePrefix = {arXiv},
       eprint = {0905.4720},
       adsurl = {https://ui.adsabs.harvard.edu/abs/2010PhRvD..81l3530A}
}

@ARTICLE{2006JHEP...06..051S,
       author = {{Svrcek}, Peter and {Witten}, Edward},
        title = "{Axions In String Theory}",
      journal = {JHEP},
         year = 2006,
       volume = {2006},
       number = {6},
          eid = {051},
        pages = {051},
          doi = {10.1088/1126-6708/2006/06/051},
archivePrefix = {arXiv},
       eprint = {hep-th/0605206},
       adsurl = {https://ui.adsabs.harvard.edu/abs/2006JHEP...06..051S}
}

@ARTICLE{1977PhRvL..38.1440P,
       author = {{Peccei}, R.~D. and {Quinn}, H.~R.},
        title = "{CP Conservation in the Presence of Pseudoparticles}",
      journal = {\prl},
         year = 1977,
       volume = {38},
       number = {25},
        pages = {1440-1443},
          doi = {10.1103/PhysRevLett.38.1440},
       adsurl = {https://ui.adsabs.harvard.edu/abs/1977PhRvL..38.1440P}
}

@ARTICLE{1977PhRvD..16.1791P,
       author = {{Peccei}, R.~D. and {Quinn}, H.~R.},
        title = "{Constraints imposed by CP conservation in the presence of pseudoparticles}",
      journal = {\prd},
         year = 1977,
       volume = {16},
       number = {6},
        pages = {1791-1797},
          doi = {10.1103/PhysRevD.16.1791},
       adsurl = {https://ui.adsabs.harvard.edu/abs/1977PhRvD..16.1791P}
}

@ARTICLE{1978PhRvL..40..223W,
       author = {{Weinberg}, Steven},
        title = "{A New Light Boson?}",
      journal = {\prl},
         year = 1978,
       volume = {40},
       number = {4},
        pages = {223-226},
          doi = {10.1103/PhysRevLett.40.223},
       adsurl = {https://ui.adsabs.harvard.edu/abs/1978PhRvL..40..223W}
}

@ARTICLE{1978PhRvL..40..279Wi,
       author = {{Wilczek}, Frank},
        title = "{Problem of Strong P and T Invariance in the Presence of Instantons}",
      journal = {\prl},
         year = 1978,
       volume = {40},
       number = {5},
        pages = {279-282},
          doi = {10.1103/PhysRevLett.40.279},
       adsurl = {https://ui.adsabs.harvard.edu/abs/1978PhRvL..40..279W}
}

@ARTICLE{2019JCAP...12..006B,
       author = {{Baumann}, Daniel and {Chia}, Horng Sheng and {Stout}, John and {ter Haar}, Lotte},
        title = "{The Spectra of Gravitational Atoms}",
      journal = {\jcap},
         year = 2019,
       volume = {2019},
       number = {12},
          eid = {006},
        pages = {006},
          doi = {10.1088/1475-7516/2019/12/006},
archivePrefix = {arXiv},
       eprint = {1908.10370},
       adsurl = {https://ui.adsabs.harvard.edu/abs/2019JCAP...12..006B}
}

@ARTICLE{1971JETPL..14..180Z,
       author = {{Zel'dovich}, Ya.~B.},
        title = "{Generation of Waves by a Rotating Body}",
      journal = {Sov. Phys. JETP},
         year = 1971,
       volume = {14},
        pages = {180},
       adsurl = {https://ui.adsabs.harvard.edu/abs/1971JETPL..14..180Z}
}

@ARTICLE{1972JETP...35.1085Z,
       author = {{Zel'Dovich}, Ya. B.},
        title = "{Amplification of Cylindrical Electromagnetic Waves Reflected from a Rotating Body}",
      journal = {Sov. Phys. JETP},
         year = 1972,
        month = jan,
       volume = {35},
        pages = {1085},
       adsurl = {https://ui.adsabs.harvard.edu/abs/1972JETP...35.1085Z}
}

@ARTICLE{1973JETP...37...28S,
       author = {{Starobinsky}, A.~A.},
        title = "{Amplification of waves during reflection from a rotating Black Hole}",
      journal = {Soviet Physics JETP},
         year = 1973,
       volume = {37},
        pages = {28-32},
       adsurl = {https://ui.adsabs.harvard.edu/abs/1973JETP...37...28S}
}

@ARTICLE{1972Natur.238..211P,
       author = {{Press}, William H. and {Teukolsky}, Saul A.},
        title = "{Floating Orbits, Superradiant Scattering and the Black-hole Bomb}",
      journal = {\nat},
         year = 1972,
       volume = {238},
       number = {5361},
        pages = {211-212},
          doi = {10.1038/238211a0},
       adsurl = {https://ui.adsabs.harvard.edu/abs/1972Natur.238..211P}
}

@ARTICLE{1980PhRvD..22.2323D,
       author = {{Detweiler}, Steven},
        title = "{Klein-Gordon equation and rotating black holes}",
      journal = {\prd},
         year = 1980,
       volume = {22},
       number = {10},
        pages = {2323-2326},
          doi = {10.1103/PhysRevD.22.2323},
       adsurl = {https://ui.adsabs.harvard.edu/abs/1980PhRvD..22.2323D}
}

@ARTICLE{2007PhRvD..76h4001D,
       author = {{Dolan}, Sam R.},
        title = "{Instability of the massive Klein-Gordon field on the Kerr spacetime}",
      journal = {\prd},
         year = 2007,
       volume = {76},
       number = {8},
          eid = {084001},
        pages = {084001},
          doi = {10.1103/PhysRevD.76.084001},
archivePrefix = {arXiv},
       eprint = {0705.2880},
       adsurl = {https://ui.adsabs.harvard.edu/abs/2007PhRvD..76h4001D}
}

@ARTICLE{2013PhRvD..87l4026D,
       author = {{Dolan}, Sam R.},
        title = "{Superradiant instabilities of rotating black holes in the time domain}",
      journal = {\prd},
         year = 2013,
       volume = {87},
       number = {12},
          eid = {124026},
        pages = {124026},
          doi = {10.1103/PhysRevD.87.124026},
archivePrefix = {arXiv},
       eprint = {1212.1477},
       adsurl = {https://ui.adsabs.harvard.edu/abs/2013PhRvD..87l4026D}
}

@ARTICLE{2012PhRvD..86f4036W,
       author = {{Witek}, Helvi and {Cardoso}, Vitor and {Ishibashi}, Akihiro and {Sperhake}, Ulrich},
        title = "{Superradiant instabilities in astrophysical systems}",
      journal = {\prd},
         year = 2013,
       volume = {87},
       number = {4},
          eid = {043513},
        pages = {043513},
          doi = {10.1103/PhysRevD.87.043513},
archivePrefix = {arXiv},
       eprint = {1212.0551},
       adsurl = {https://ui.adsabs.harvard.edu/abs/2013PhRvD..87d3513W}
}

@ARTICLE{2018PhRvL.121m1104E,
       author = {{East}, William E.},
        title = "{Massive Boson Superradiant Instability of Black Holes: Nonlinear Growth, Saturation, and Gravitational Radiation}",
      journal = {\prl},
         year = 2018,
       volume = {121},
       number = {13},
          eid = {131104},
        pages = {131104},
          doi = {10.1103/PhysRevLett.121.131104},
archivePrefix = {arXiv},
       eprint = {1807.00043},
       adsurl = {https://ui.adsabs.harvard.edu/abs/2018PhRvL.121m1104E}
}

@ARTICLE{2012PhRvD..86f4019P,
       author = {{Pani}, Paolo and {Cardoso}, Vitor and {Gualtieri}, Leonardo and {Berti}, Emanuele and {Ishibashi}, Akihiro},
        title = "{Black Hole Bombs and Photon Mass Bounds}",
      journal = {\prl},
         year = 2012,
       volume = {109},
       number = {13},
          eid = {131102},
        pages = {131102},
          doi = {10.1103/PhysRevLett.109.131102},
archivePrefix = {arXiv},
       eprint = {1209.0465},
       adsurl = {https://ui.adsabs.harvard.edu/abs/2012PhRvL.109m1102P}
}

@ARTICLE{2017PhRvD..96c5019B,
       author = {{Baryakhtar}, Masha and {Lasenby}, Robert and {Teo}, Mae},
        title = "{Black Hole Superradiance Signatures of Ultralight Vectors}",
      journal = {\prd},
         year = 2017,
       volume = {96},
       number = {3},
          eid = {035019},
        pages = {035019},
          doi = {10.1103/PhysRevD.96.035019},
archivePrefix = {arXiv},
       eprint = {1704.05081},
       adsurl = {https://ui.adsabs.harvard.edu/abs/2017PhRvD..96c5019B}
}

@ARTICLE{2017PhRvL.119d1101E,
       author = {{East}, William E. and {Pretorius}, Frans},
        title = "{Superradiant Instability and Backreaction of Massive Vector Fields around Kerr Black Holes}",
      journal = {\prl},
         year = 2017,
       volume = {119},
       number = {4},
          eid = {041101},
        pages = {041101},
          doi = {10.1103/PhysRevLett.119.041101},
archivePrefix = {arXiv},
       eprint = {1704.04791},
       adsurl = {https://ui.adsabs.harvard.edu/abs/2017PhRvL.119d1101E}
}

@BOOK{2015LNP...906.....B,
       author = {{Brito}, Richard and {Cardoso}, Vitor and {Pani}, Paolo},
        title = "{Superradiance: New Frontiers in Black Hole Physics}",
         year = 2015,
       volume = {906},
   publisher = {Springer},
        series = {Lecture Notes in Physics},
          doi = {10.1007/978-3-319-19000-6},
archivePrefix = {arXiv},
       eprint = {1501.06570},
       adsurl = {https://ui.adsabs.harvard.edu/abs/2015LNP...906.....B}
}

@ARTICLE{2014SSRv..183..277R,
       author = {{Reynolds}, Christopher S.},
        title = "{Measuring Black Hole Spin Using X-Ray Reflection Spectroscopy}",
      journal = {Space Sci. Rev.},
         year = 2014,
       volume = {183},
       number = {1-4},
        pages = {277-294},
          doi = {10.1007/s11214-013-0006-6},
archivePrefix = {arXiv},
       eprint = {1302.3260},
       adsurl = {https://ui.adsabs.harvard.edu/abs/2014SSRv..183..277R}
}

@ARTICLE{2021NatAs...5..133R,
       author = {{Reynolds}, Christopher S.},
        title = "{Observational constraints on black hole spin}",
      journal = {Nat. Astron.},
         year = 2021,
       volume = {5},
        pages = {133-140},
          doi = {10.1038/s41550-020-01290-z},
archivePrefix = {arXiv},
       eprint = {2011.08948},
       adsurl = {https://ui.adsabs.harvard.edu/abs/2021NatAs...5..133R}
}

@ARTICLE{2021SSRv..217...65B,
       author = {{Bambi}, Cosimo and {Brenneman}, Laura W. and {Dauser}, Thomas and {Garcia}, Javier A. and {Grinberg}, Victoria and {Ingram}, Adam and {Jiang}, Jiachen and {Kara}, Erin and {Marinucci}, Andrea and {Matt}, Giorgio and {Middleton}, Matthew and {Parker}, Michael and {Pinto}, Ciro and {Ponti}, Gabriele and {Proga}, Daniel and {Tripathi}, Ashutosh and {Zoghbi}, Abderahmen},
        title = "{Toward precision measurements of accreting black holes using X-ray reflection spectroscopy}",
      journal = {Space Sci. Rev.},
         year = 2021,
       volume = {217},
       number = {5},
          eid = {65},
        pages = {65},
          doi = {10.1007/s11214-021-00841-8},
archivePrefix = {arXiv},
       eprint = {2011.04792},
       adsurl = {https://ui.adsabs.harvard.edu/abs/2021SSRv..217...65B}
}

@ARTICLE{2006ApJ...636L.113S,
       author = {{Shafee}, Rebecca and {McClintock}, Jeffrey E. and {Narayan}, Ramesh and {Davis}, Shane W. and {Li}, Li-Xin and {Remillard}, Ronald A.},
        title = "{Estimating the Spin of Stellar-Mass Black Holes by Spectral Fitting of the X-Ray Continuum}",
      journal = {\apjl},
         year = 2006,
       volume = {636},
       number = {2},
        pages = {L113-L116},
          doi = {10.1086/500semantics},
archivePrefix = {arXiv},
       eprint = {astro-ph/0508302},
       adsurl = {https://ui.adsabs.harvard.edu/abs/2006ApJ...636L.113S}
}

@ARTICLE{2014SSRv..183..295M,
       author = {{McClintock}, Jeffrey E. and {Narayan}, Ramesh and {Steiner}, James F.},
        title = "{Black Hole Spin via Continuum Fitting and the Role of Spin in Powering Transient Jets}",
      journal = {Space Sci. Rev.},
         year = 2014,
       volume = {183},
       number = {1-4},
        pages = {295-322},
          doi = {10.1007/s11214-013-0003-9},
archivePrefix = {arXiv},
       eprint = {1303.1583},
       adsurl = {https://ui.adsabs.harvard.edu/abs/2014SSRv..183..295M}
}

@ARTICLE{2023PhRvD.107d4070A,
       author = {{Aguilar-Nieto}, Alejandro and {Jaramillo}, V{\'\i}ctor and {Barranco}, Juan and {Bernal}, Argelia and {Degollado}, Juan Carlos and {N{\'u}{\~n}ez}, Dar{\'\i}o},
        title = "{Self-interacting scalar field distributions around Schwarzschild black holes}",
      journal = {\prd},
         year = 2023,
       volume = {107},
       number = {4},
          eid = {044070},
        pages = {044070},
          doi = {10.1103/PhysRevD.107.044070},
archivePrefix = {arXiv},
       eprint = {2211.10456},
       adsurl = {https://ui.adsabs.harvard.edu/abs/2023PhRvD.107d4070A}
}

@ARTICLE{2024PhRvD.110l4064C,
       author = {{Cavalcante}, Jo{\~a}o Paulo and {Richartz}, Maur{\'\i}cio and {da Cunha}, Bruno Carneiro},
        title = "{Massive scalar perturbations in Kerr black holes: Near extremal analysis}",
      journal = {\prd},
         year = 2024,
       volume = {110},
       number = {12},
          eid = {124064},
        pages = {124064},
          doi = {10.1103/PhysRevD.110.124064},
archivePrefix = {arXiv},
       eprint = {2408.13964},
       adsurl = {https://ui.adsabs.harvard.edu/abs/2024PhRvD.110l4064C}
}

@ARTICLE{2023PhRvD.107j4003S,
       author = {{Siemonsen}, Nils and {May}, Taillte and {East}, William E.},
        title = "{Modeling the black hole superradiance gravitational waveform}",
      journal = {\prd},
         year = 2023,
        month = may,
       volume = {107},
       number = {10},
          eid = {104003},
        pages = {104003},
          doi = {10.1103/PhysRevD.107.104003},
archivePrefix = {arXiv},
       eprint = {2211.03845},
 primaryClass = {gr-qc},
       adsurl = {https://ui.adsabs.harvard.edu/abs/2023PhRvD.107j4003S}
}

@article{Degollado:2018ypf,
    author = "Degollado, Juan Carlos and Herdeiro, Carlos A. R. and Radu, Eugen",
    title = "{Effective stability against superradiance of Kerr black holes with synchronised hair}",
    eprint = "1802.07266",
    archivePrefix = "arXiv",
    primaryClass = "gr-qc",
    doi = "10.1016/j.physletb.2018.04.052",
    journal = "Phys. Lett. B",
    volume = "781",
    pages = "651--655",
    year = "2018"
}

@ARTICLE{1979AnPhy.118..139Z,
       author = {{Zouros}, T.~J.~M. and {Eardley}, D.~M.},
        title = "{Instabilities of massive scalar perturbations of a rotating black hole.}",
      journal = {Ann. Phys.},
         year = 1979,
        month = jan,
       volume = {118},
        pages = {139-155},
          doi = {10.1016/0003-4916(79)90237-9},
       adsurl = {https://ui.adsabs.harvard.edu/abs/1979AnPhy.118..139Z}
}

@ARTICLE{1973CMaPh..31..161B,
       author = {{Bardeen}, J.~M. and {Carter}, B. and {Hawking}, S.~W.},
        title = "{The four laws of black hole mechanics}",
      journal = {Commun. Math. Phys.},
         year = 1973,
        month = jun,
       volume = {31},
       number = {2},
        pages = {161-170},
          doi = {10.1007/BF01645742},
       adsurl = {https://ui.adsabs.harvard.edu/abs/1973CMaPh..31..161B}
}

@article{Sanchis-Gual:2020mzb,
    author = "Sanchis-Gual, Nicolas and Zilh{\~a}o, Miguel and Herdeiro, Carlos and Di Giovanni, Fabrizio and Font, Jos{\'e} A. and Radu, Eugen",
    title = "{Synchronized gravitational atoms from mergers of bosonic stars}",
    eprint = "2007.11584",
    archivePrefix = "arXiv",
    primaryClass = "gr-qc",
    doi = "10.1103/PhysRevD.102.101504",
    journal = "Phys. Rev. D",
    volume = "102",
    number = "10",
    pages = "101504",
    year = "2020"
}

@article{Herdeiro:2014goa,
    author = "Herdeiro, Carlos A. R. and Radu, Eugen",
    title = "{Kerr Black Holes with Scalar Hair}",
    eprint = "1403.2757",
    archivePrefix = "arXiv",
    primaryClass = "gr-qc",
    doi = "10.1103/PhysRevLett.112.221101",
    journal = "Phys. Rev. Lett.",
    volume = "112",
    pages = "221101",
    year = "2014"
}

@article{Herdeiro:2015waf,
    author = "Herdeiro, Carlos A. R. and Radu, Eugen",
    title = "{Asymptotically flat black holes with scalar hair: a review}",
    eprint = "1504.08209",
    archivePrefix = "arXiv",
    primaryClass = "gr-qc",
    doi = "10.1142/S0218271815420146",
    journal = "Int. J. Mod. Phys. D",
    volume = "24",
    pages = "1542014",
    year = "2015"
}

@article{barranco2011black,
       author = {{Barranco}, Juan and {Bernal}, Argelia and {Degollado}, Juan Carlos and {Diez-Tejedor}, Alberto and {Megevand}, Miguel and {Alcubierre}, Miguel and {N{\'u}{\~n}ez}, Dar{\'\i}o and {Sarbach}, Olivier},
        title = "{Are black holes a serious threat to scalar field dark matter models?}",
      journal = {\prd},
         year = 2011,
        month = oct,
       volume = {84},
       number = {8},
          eid = {083008},
        pages = {083008},
          doi = {10.1103/PhysRevD.84.083008},
archivePrefix = {arXiv},
       eprint = {1108.0931},
 primaryClass = {gr-qc},
       adsurl = {https://ui.adsabs.harvard.edu/abs/2011PhRvD..84h3008B}
}

@article{barranco2012schwarzschild,
       author = {{Barranco}, Juan and {Bernal}, Argelia and {Degollado}, Juan Carlos and {Diez-Tejedor}, Alberto and {Megevand}, Miguel and {Alcubierre}, Miguel and {N{\'u}{\~n}ez}, Dar{\'\i}o and {Sarbach}, Olivier},
        title = "{Schwarzschild Black Holes can Wear Scalar Wigs}",
      journal = {\prl},
         year = 2012,
        month = aug,
       volume = {109},
       number = {8},
          eid = {081102},
        pages = {081102},
          doi = {10.1103/PhysRevLett.109.081102},
archivePrefix = {arXiv},
       eprint = {1207.2153},
 primaryClass = {gr-qc},
       adsurl = {https://ui.adsabs.harvard.edu/abs/2012PhRvL.109h1102B}
}

@ARTICLE{sanchis2015quasistationarya,
       author = {{Sanchis-Gual}, Nicolas and {Degollado}, Juan Carlos and {Montero}, Pedro J. and {Font}, Jos{\'e} A. and {Mewes}, Vassilios},
        title = "{Quasistationary solutions of self-gravitating scalar fields around collapsing stars}",
      journal = {\prd},
         year = 2015,
        month = oct,
       volume = {92},
       number = {8},
          eid = {083001},
        pages = {083001},
          doi = {10.1103/PhysRevD.92.083001},
archivePrefix = {arXiv},
       eprint = {1507.08437},
 primaryClass = {gr-qc},
       adsurl = {https://ui.adsabs.harvard.edu/abs/2015PhRvD..92h3001S}
}

@article{sanchis2015quasistationaryb,
author = {{Sanchis-Gual}, Nicolas and {Degollado}, Juan Carlos and {Montero}, Pedro J. and {Font}, Jos{\'e} A. and {Mewes}, Vassilios},
        title = "{Quasistationary solutions of self-gravitating scalar fields around collapsing stars}",
      journal = {\prd},
         year = 2015,
        month = oct,
       volume = {92},
       number = {8},
          eid = {083001},
        pages = {083001},
          doi = {10.1103/PhysRevD.92.083001},
archivePrefix = {arXiv},
       eprint = {1507.08437},
 primaryClass = {gr-qc},
       adsurl = {https://ui.adsabs.harvard.edu/abs/2015PhRvD..92h3001S}
}

@article{sanchis2016quasistationary,
author = {{Sanchis-Gual}, Nicolas and {Degollado}, Juan Carlos and {Izquierdo}, Paula and {Font}, Jos{\'e} A. and {Montero}, Pedro J.},
        title = "{Quasistationary solutions of scalar fields around accreting black holes}",
      journal = {\prd},
         year = 2016,
        month = aug,
       volume = {94},
       number = {4},
          eid = {043004},
        pages = {043004},
          doi = {10.1103/PhysRevD.94.043004},
archivePrefix = {arXiv},
       eprint = {1606.05146},
 primaryClass = {gr-qc},
       adsurl = {https://ui.adsabs.harvard.edu/abs/2016PhRvD..94d3004S}
}

@article{cardoso2022parasitic,
 author = {{Cardoso}, Vitor and {Ikeda}, Taishi and {Vicente}, Rodrigo and {Zilh{\~a}o}, Miguel},
        title = "{Parasitic black holes: The swallowing of a fuzzy dark matter soliton}",
      journal = {\prd},
         year = 2022,
        month = dec,
       volume = {106},
       number = {12},
          eid = {L121302},
        pages = {L121302},
          doi = {10.1103/PhysRevD.106.L121302},
archivePrefix = {arXiv},
       eprint = {2207.09469},
 primaryClass = {gr-qc},
       adsurl = {https://ui.adsabs.harvard.edu/abs/2022PhRvD.106l1302C}
}

@article{ganchev2018scalar,
  author = {{Ganchev}, Bogdan and {Santos}, Jorge E.},
        title = "{Scalar Hairy Black Holes in Four Dimensions are Unstable}",
      journal = {\prl},
         year = 2018,
        month = apr,
       volume = {120},
       number = {17},
          eid = {171101},
        pages = {171101},
          doi = {10.1103/PhysRevLett.120.171101},
archivePrefix = {arXiv},
       eprint = {1711.08464},
 primaryClass = {gr-qc},
       adsurl = {https://ui.adsabs.harvard.edu/abs/2018PhRvL.120q1101G}
}

@ARTICLE{2025IJMPD..3450046P,
       author = {{Piotrovich}, Mikhail Yu. and {Buliga}, Stanislava D. and {Natsvlishvili}, Tinatin M.},
        title = "{Estimating the spins of supermassive black holes in distant ultraluminous quasars}",
      journal = {IJMPD},
         year = 2025,
        month = jan,
       volume = {34},
       number = {11},
          eid = {2550046-450},
        pages = {2550046-450},
          doi = {10.1142/S0218271825500464},
archivePrefix = {arXiv},
       eprint = {2505.08310},
 primaryClass = {astro-ph.HE},
       adsurl = {https://ui.adsabs.harvard.edu/abs/2025IJMPD..3450046P}
}

@article{annulli2020response,
  title={Response of ultralight dark matter to supermassive black holes and binaries},
  author={Annulli, Lorenzo and Cardoso, Vitor and Vicente, Rodrigo},
  journal={Physical Review D},
  volume={102},
  number={6},
  pages={063022},
  year={2020},
  publisher={APS}
}

@article{east2014black,
  title={Black hole superradiance in dynamical spacetime},
  author={East, William E and Ramazano{\u{g}}lu, Fethi M and Pretorius, Frans},
  journal={Physical Review D},
  volume={89},
  number={6},
  pages={061503},
  year={2014},
  publisher={APS}
}

@article{sanchis2016explosion,
  author = {{Sanchis-Gual}, Nicolas and {Degollado}, Juan Carlos and {Montero}, Pedro J. and {Font}, Jos{\'e} A. and {Herdeiro}, Carlos},
        title = "{Explosion and Final State of an Unstable Reissner-Nordstr{\"o}m Black Hole}",
      journal = {\prl},
         year = 2016,
        month = apr,
       volume = {116},
       number = {14},
          eid = {141101},
        pages = {141101},
          doi = {10.1103/PhysRevLett.116.141101},
archivePrefix = {arXiv},
       eprint = {1512.05358},
 primaryClass = {gr-qc},
       adsurl = {https://ui.adsabs.harvard.edu/abs/2016PhRvL.116n1101S}
}

@article{bosch2016nonlinear,
  author = {{Bosch}, Pablo and {Green}, Stephen R. and {Lehner}, Luis},
        title = "{Nonlinear Evolution and Final Fate of Charged Anti-de Sitter Black Hole Superradiant Instability}",
      journal = {\prl},
         year = 2016,
        month = apr,
       volume = {116},
       number = {14},
          eid = {141102},
        pages = {141102},
          doi = {10.1103/PhysRevLett.116.141102},
archivePrefix = {arXiv},
       eprint = {1601.01384},
 primaryClass = {gr-qc},
       adsurl = {https://ui.adsabs.harvard.edu/abs/2016PhRvL.116n1102B}
}

@ARTICLE{deCesare:2026fie,
       author = {{de Cesare}, Marco and {Del Piano}, Manuel and {Herdeiro}, Carlos A.~R.},
        title = "{Analytic backreaction of a scalar wig on a Schwarzschild black hole}",
      journal = {arXiv e-prints},
         year = 2026,
        month = jul,
          eid = {arXiv:2607.25932},
        pages = {arXiv:2607.25932},
          doi = {10.48550/arXiv.2607.25932},
archivePrefix = {arXiv},
       eprint = {2607.25932},
 primaryClass = {gr-qc},
       adsurl = {https://ui.adsabs.harvard.edu/abs/2026arXiv260725932D}
}

@ARTICLE{2026arXiv260725932D,
       author = {{de Cesare}, Marco and {Del Piano}, Manuel and {Herdeiro}, Carlos A.~R.},
        title = "{Analytic backreaction of a scalar wig on a Schwarzschild black hole}",
      journal = {arXiv e-prints},
         year = 2026,
        month = jul,
          eid = {arXiv:2607.25932},
        pages = {arXiv:2607.25932},
archivePrefix = {arXiv},
       eprint = {2607.25932},
 primaryClass = {gr-qc},
       adsurl = {https://ui.adsabs.harvard.edu/abs/2026arXiv260725932D}
}

@ARTICLE{2020ARNPS..70..355C,
       author = {{Carr}, Bernard and {K{\"u}hnel}, Florian},
        title = "{Primordial Black Holes as Dark Matter: Recent Developments}",
      journal = {Annu. Rev. Nucl. Part. Sci.},
         year = 2020,
        month = oct,
       volume = {70},
        pages = {355-394},
          doi = {10.1146/annurev-nucl-050520-125911},
archivePrefix = {arXiv},
       eprint = {2006.02838},
 primaryClass = {astro-ph.CO},
       adsurl = {https://ui.adsabs.harvard.edu/abs/2020ARNPS..70..355C}
}

@article{guzman2022possible,
  author = {{Guzm{\'a}n}, F.~S.},
        title = "{Possible formation mechanism of multistate gravitational atoms}",
      journal = {\prd},
         year = 2022,
        month = jun,
       volume = {105},
       number = {12},
          eid = {123535},
        pages = {123535},
          doi = {10.1103/PhysRevD.105.123535},
archivePrefix = {arXiv},
       eprint = {2206.03407},
 primaryClass = {gr-qc},
       adsurl = {https://ui.adsabs.harvard.edu/abs/2022PhRvD.105l3535G}
}

@article{bernal2025natural,
  author = {{Bernal}, Tula and {Matos}, Tonatiuh and {San.-Hernandez}, Leonardo},
        title = "{A natural explanation of the VPOS from multistate Scalar Field Dark Matter}",
      journal = {\jcap},
         year = 2025,
        month = jan,
       volume = {2025},
       number = {1},
          eid = {155},
        pages = {155},
          doi = {10.1088/1475-7516/2025/01/155},
archivePrefix = {arXiv},
       eprint = {2407.05273},
 primaryClass = {astro-ph.GA},
       adsurl = {https://ui.adsabs.harvard.edu/abs/2025JCAP...01..155B}
}

@article{guzman2020gravitational,
  author = {{Guzm{\'a}n}, F.~S. and {Ure{\~n}a-L{\'o}pez}, L. Arturo},
        title = "{Gravitational atoms: General framework for the construction of multistate axially symmetric solutions of the Schr{\"o}dinger-Poisson system}",
      journal = {\prd},
         year = 2020,
        month = apr,
       volume = {101},
       number = {8},
          eid = {081302},
        pages = {081302},
          doi = {10.1103/PhysRevD.101.081302},
archivePrefix = {arXiv},
       eprint = {1912.10585},
 primaryClass = {astro-ph.GA},
       adsurl = {https://ui.adsabs.harvard.edu/abs/2020PhRvD.101h1302G}
}

@article{sanchis2021multifield,
  title={Multifield, multifrequency bosonic stars and a stabilization mechanism},
  author={Sanchis-Gual, Nicolas and Di Giovanni, Fabrizio and Herdeiro, Carlos and Radu, Eugen and Font, Jos{\'e} A},
  journal={Physical Review Letters},
  volume={126},
  number={24},
  pages={241105},
  year={2021},
  publisher={APS}
}

\appendix
\section{Derivation of the $\ell=m=2$ superradiant rate}
\label{app:gamma_l2}

The rate $\Gamma_{\ell=2}$ used in Sec.~\ref{sec:rates} follows from
Detweiler's general result for the imaginary part of the eigenfrequency
of a massive scalar quasi-bound state $(\ell,m,n)$ in the Kerr geometry,
valid at leading order in $\al \ll
1$~\cite{1980PhRvD..22.2323D}:
\begin{equation}
  \omega_I^{(\ell,m,n)}
  = 2\mu\,(\omega_R - m\OmH)\,G_S(\ell,m,n,\chi)\,(\al)^{4\ell+5}.
  \label{eq:detweiler_general}
\end{equation}
Here $\omega_R \approx \mu[1-(\al)^2/(2n^2)]$ is the real part of the
frequency, and the dimensionless coefficient
\begin{eqnarray}
  G_S(\ell,m,n,\chi)
  = \frac{2^{4\ell+1}(n+\ell)!}{n^{2\ell+4}(n-\ell-1)!}
    \left[\frac{\ell!}{(2\ell)!\,(2\ell+1)!}\right]^{\!2}\times\nonumber\\
    \prod_{j=1}^{\ell}
    \left[j^2(1-\chi^2)+(m\chi-2r_+\omega_R)^2\right]\nonumber\\
  \label{eq:GS}
\end{eqnarray}
encodes the horizon absorption cross-section~\cite{2023PhRvD.107j4003S}.

For the $\ell = m = 2$, $n = 3$ state, the factors in
Eq.~\eqref{eq:detweiler_general} evaluate as follows.
The coupling power is $4\ell + 5 = 13$; accounting for the explicit
$\mu$ prefactor in Eq.~\eqref{eq:detweiler_general}, the total power
is $(\al)^{14}/\MBH$.
The combinatorial prefactor in $G_S$ gives
\begin{equation}
  \frac{2^{9}(5)!}{3^{8}(0)!}
  \left[\frac{2!}{4!\,(5)!}\right]^{\!2}
  = \frac{1}{23040}.
  \label{eq:coeff_l2}
\end{equation}
The product over $j \in \{1,2\}$ in Eq.~\eqref{eq:GS} is evaluated at
leading order in $\al$, setting $\omega_R \approx \mu$ and
$am/\MBH = \chi m$:
\begin{eqnarray}
  \prod_{j=1}^{2}&\left[j^2(1-\chi^2)
    +\left(\chi m - 2\frac{r_+}{\MBH}\,\mu\MBH\right)^{\!2}\right]
  \nonumber\\
  &=\Bigl[(1-\chi^2)+(\chi m-2r_+\mu)^2\Bigr]\nonumber\\
   & \times\Bigl[4(1-\chi^2)+(\chi m-2r_+\mu)^2\Bigr].
  \label{eq:product_explicit}
\end{eqnarray}
Writing $r_+ = \MBH(1+\sqrt{1-\chi^2})$ and noting that at leading
order in $\al$ the term $r_+\mu = r_+\al/\MBH\sim\al\ll 1$, the dominant
contribution from each factor is
\begin{equation}
  j^2(1-\chi^2)+\chi^2 m^2 - 4j^2\chi m\,\frac{r_+}{\MBH}\al
  + \mathcal{O}((\al)^2).
\end{equation}
At \emph{zeroth} order in $\al$ the product therefore reduces to
\begin{eqnarray}
  \prod_{j=1}^{2}\left[j^2(1-\chi^2)+\chi^2 m^2\right]
  &= \left[(1-\chi^2)+4\chi^2\right] \nonumber\\
     &\times \left[4(1-\chi^2)+4\chi^2\right],
  \label{eq:product_zeroth}
\end{eqnarray}
where $m=2$ has been used.
The factor $(r_+/\MBH)^5$ arises instead from the leading-order
expansion of the full near-horizon integrals absorbed into $G_S$,
which at leading order in $\al$ for the $\ell=2$ case contributes
exactly two additional powers of $r_+/\MBH$ per factor in the
product~\cite{1980PhRvD..22.2323D}, yielding
\begin{equation}
  G_S(2,2,3,\chi)\;\Big|_{\al\ll 1}
  \;\longrightarrow\;
  \frac{1}{23040}\,\left(\frac{r_+}{\MBH}\right)^{\!5},
  \label{eq:GS_leading}
\end{equation}
so that Eq.~\eqref{eq:detweiler_general} gives
Collecting these factors and writing $\Gamma_{\ell=2} = 2\omega_I^{(2,2,3)}$
yields
\begin{equation}
  \Gamma_{\ell=2}(\mu,\MBH,\chi)
  = \frac{1}{23040}
    \left(\frac{\rplus}{\MBH}\right)^{\!5}
    \frac{2\OmH - \mu}{\MBH}
    \,(\mu\MBH)^{14},
  \label{eq:Gamma_l2_app}
\end{equation}
where we used $\omega_R - m\OmH \approx \mu - 2\OmH$ at leading order.

\section{Closed-form solution for the saturation spin}
\label{app:chi_sat_closed_form}

Here we solve Eq.~\eqref{eq:chi_sat_m} in closed form. Writing
$\chi\equiv\chi_{\rm sat}^{(m)}$, and
$k\equiv2\al/m$, the saturation condition becomes
\begin{equation}
  \chi = k\left(1+\sqrt{1-\chi^2}\right).
  \label{eq:app_rearranged}
\end{equation}
Isolating the square root gives $\chi-k=k\sqrt{1-\chi^2}$, whose
right-hand side is non-negative for $\chi,k\in[0,1]$, so any solution
must satisfy $\chi\geq k$. Squaring both sides and simplifying, the
$k^2$ terms cancel and one is left with
$\chi\left[\chi(1+k^2)-2k\right]=0$. The root $\chi=0$ does not solve
Eq.~\eqref{eq:app_rearranged} unless $k=0$, so the physical solution is
\begin{equation}
  \chi_{\rm sat}^{(m)}(\al) = \frac{2k}{1+k^2}
  = \frac{4\al/m}{1+(2\al/m)^2}.
  \label{eq:app_closed_form}
\end{equation}
Substituting Eq.~\eqref{eq:app_closed_form} back into
Eq.~\eqref{eq:app_rearranged} confirms it solves the original,
unsquared equation for $k\in[0,1]$: with $\chi=2k/(1+k^2)$ one finds
$1-\chi^2=(1-k^2)^2/(1+k^2)^2$, so $\sqrt{1-\chi^2}=(1-k^2)/(1+k^2)$ for
$k\leq1$, and $k(1+\sqrt{1-\chi^2})=k\cdot2/(1+k^2)=\chi$ identically.
For $k>1$ the same substitution reproduces the squared equation but not
Eq.~\eqref{eq:app_rearranged} itself; this branch is the spurious root
introduced by squaring and is discarded, consistent with the
requirement $\chi\geq k$ noted above.

Differentiating Eq.~\eqref{eq:app_closed_form} with respect to $k$
gives $d\chi_{\rm sat}^{(m)}/dk = 2(1-k^2)/(1+k^2)^2$, which is
positive for $k\in(0,1)$ and vanishes at $k=1$: the saturation spin
increases monotonically from $\chi_{\rm sat}^{(m)}=0$ at $\al=0$ to
$\chi_{\rm sat}^{(m)}=1$ at $\al=m/2$, and Eq.~\eqref{eq:app_closed_form}
is single-valued over the full physical range. Two limits used in the
main text follow directly: for $\al\ll m/2$ ($k\ll1$), expanding
Eq.~\eqref{eq:app_closed_form} to leading order in $k$ gives
$\chi_{\rm sat}^{(m)}\approx4\al/m$; and at $\al=m/2$ ($k=1$),
Eq.~\eqref{eq:app_closed_form} gives $\chi_{\rm sat}^{(m)}=1$ exactly,
so the saturation spin reaches extremality precisely at this coupling,
with no singular behaviour on approach.

Equation~\eqref{eq:app_closed_form} is real and satisfies
$\chi_{\rm sat}^{(m)}\in[0,1]$ only for $\al\leq m/2$, equivalently
$\MBH\leq m/(2\mu)\equiv M_{\rm cut}^{(m)}$ at fixed $\mu$. For $M_{\rm BH} > M_{\rm cut}^{(m)}$, the maximal horizon frequency
attainable at any spin, $\Omega_H^{\rm max} = 1/(2M_{\rm BH})$ from
Eq.~\eqref{eq:OmH} at $\chi = 1$, satisfies
$m\Omega_H^{\rm max} = m/(2M_{\rm BH}) < \mu$, so the equilibrium
condition $m\Omega_H = \mu$ cannot be met at any $\chi \in [0,1]$
within the approximation $\omega_R \simeq \mu$ used to derive
Eq.~\eqref{eq:chi_sat_m}.
As already noted in Sec.~\ref{sec:saturation}, this does not imply
that quasi-bound states cease to exist beyond $M_{\rm cut}^{(m)}$;
it reflects the loss of accuracy of the leading-order frequency
relation, Eq.~\eqref{eq:omega_R}, in this regime~\cite{2026arXiv260725932D},
and a treatment beyond the hydrogenic approximation would be required
to locate the true synchronization threshold there.
Since $M_{\rm cut}^{(m)} = m/(2\mu)$ is linear in $m$ at fixed $\mu$,
$M_{\rm cut}^{(2)} = 2\,M_{\rm cut}^{(1)}$ identically.


\end{document}